\documentclass[preprint2]{aastex}
\newcommand{\AlIII}{Al~{\sc iii}}
\newcommand{\CIV}{C~{\sc iv}}
\newcommand{\CaII}{Ca~{\sc ii}}
\newcommand{\Ha}{H$\alpha$}
\newcommand{\Hb}{H$\beta$}

\newcommand{\HeI}{He~{\sc i}}
\newcommand{\FeII}{Fe~{\sc ii}}
\newcommand{\feka}{\hbox{Fe~K$\alpha$}}

\newcommand{\MgII}{Mg~{\sc ii}}

\newcommand{\NV}{N~{\sc v}}

\newcommand{\SIV}{S~{\sc iv}}
\newcommand{\OIII}{[O~{\sc iii}]}
\newcommand{\kms}{\hbox{km~s$^{-1}$}}
\newcommand{\cmsq}{\hbox{cm$^{-2}$}}
\newcommand{\cc}{\hbox{cm$^{-3}$}}
\newcommand{\flux}{\hbox{erg~cm$^{-2}$~s$^{-1}$}}
\newcommand{\lumin}{\hbox{erg~s$^{-1}$}}
\newcommand{\persec}{\hbox{s$^{-1}$}}
\newcommand{\msun}{\hbox{${M}_{\odot}$}}
\newcommand{\lsun}{\hbox{${L}_{\odot}$}}
\newcommand{\nh}{\hbox{${N}_{\rm H}$}} 
\newcommand{\nim}{Nuclear Instruments \& Methods}
\newcommand{\mrk}{Mrk~231}
\newcommand{\asca}{{\emph{ASCA}}}
\newcommand{\chandra}{{\emph{Chandra}}}
\newcommand{\sax}{{\emph{Beppo-SAX}}}
\newcommand{\rosat}{{\emph{ROSAT}}}
\newcommand{\einstein}{{\emph{Einstein}}}
\newcommand{\heao}{\hbox{\emph{HEAO-1}}}
\newcommand{\hst}{{\emph{HST}}}
\newcommand{\iue}{{\emph{IUE}}}
\newcommand{\iras}{{\emph{IRAS}}}

\newcommand{\aox}{$\alpha_{\rm ox}$}

\begin{document}
 
%\received{date month year}
%\accepted{date month year}
%\journalid{number}{date month year}
%\articleid{number}{number}
%\slugcomment{submitted to: {\it The Astrophysical Journal Letters}}
 
\shortauthors{Gallagher et al.}
\shorttitle{X-raying Mrk~231 with \chandra}

%%%%%%%%%%%%%%%%%%%%%%%%%%%%%%%%%%%%%%%%%%%%%%%%%%%%%%%%%%%%%%%%%%%%%%%%%%%%%%%%%%

\title{X-raying the Ultraluminous Infrared Starburst Galaxy and Broad
Absorption Line QSO, Markarian 231, with \chandra.}

\author{S.~C.~Gallagher, W.~N.~Brandt, G.~Chartas, G.~P.~Garmire,}
       
\affil{Department of Astronomy and Astrophysics \\ 
       The Pennsylvania State University \\
       University Park, PA 16802 \\ 
       {\em gallsc, niel, chartas, garmire@astro.psu.edu}}

\and
\author{R.~M.~Sambruna}
\affil{Department of Physics \& Astronomy and School of Computational Sciences\\
	George Mason University \\
        4400 University Drive M/S 3F3\\
	Fairfax, VA 22030-4444\\
      {\em rms@physics.gmu.edu}}

\begin{abstract}
With 40~ks of \chandra\ ACIS-S3 exposure, new
information on both the starburst and QSO components of the X-ray
emission of Markarian 231, an ultraluminous infrared galaxy and Broad Absorption
Line QSO, has been obtained.  The bulk of the X-ray luminosity is
emitted from an unresolved nuclear point source, and the spectrum is
remarkably hard with the majority of the flux emitted above 2~keV.  
Most notably, significant nuclear variability 
(a decrease of $\sim45\%$ in approximately 6 hours) at
energies above 2~keV indicates that \chandra\ has probed within 
light hours of the central black hole.  Though we concur with Maloney
\& Reynolds that the direct continuum is not observed, 
this variability coupled with the 188~eV upper limit on the equivalent width
of the \feka\ emission line 
argues against the reflection-dominated model put forth by these authors
based on their \asca\ data.  Instead, we favor a model in
which a small, Compton-thick absorber blocks the direct X-rays, and
only indirect, scattered X-rays from  multiple lines of sight can reach the observer.
Extended soft, thermal emission
encompasses the optical extent of the galaxy and
exhibits resolved structure.  An off-nuclear X-ray source with a
0.35--8.0~keV luminosity of $L_{\rm X}=7\times10^{39}$~erg~s$^{-1}$,
consistent with the ultraluminous X-ray sources in other nearby
starbursts, is detected.
We also present an unpublished FOS spectrum from the \hst\ archive
showing the broad \CIV\ absorption. 
\end{abstract}

%%%%%%%%%%%%%%%%%%%%%%%%%%%%%%%%%%%%%%%%%%%%%%%%%%%%%%%%%%%%%%%%%%%%%%%%%%%%%%%%%%

\section{Introduction}
\label{sec:intro}

The extraordinary galaxy Markarian~231 was discovered in 1969 as part of a
survey searching for galaxies with strong ultraviolet continua \citep{Ma1969}.
With the first spectrum, it was identified as unique among the
Markarian sample because of deep, blue-shifted Na~D absorption evocative
of broad absorption line quasi-stellar objects \citep[BAL~QSOs;][]{ArDiEs1971}.
Soon after, \citet{AdWe1972} also identified broad
absorption lines from \CaII\ H and K and \HeI\ at the same velocity as the Na~D system.
In addition, they first described the long tidal tails and disturbed morphology of the 
galaxy.

Since those early observations, \mrk\ has maintained its reputation as an exceptional 
object and continues to be a favorite target in all wavelength regimes.  
Years before the {\it Infrared Astronomical Satellite} (\iras) 
was launched, \mrk\ was known to have an
infrared luminosity on a par with QSOs 
\citep[$\sim4\times10^{12}$~\lsun;][]{RiLo1972,RiLo1975}.
\citet{We1973} initially proposed that this energy could be non-thermal ultraviolet 
emission reradiated thermally by dust.  Even after \iras\ expanded the known population
of ultraluminous infrared galaxies, \mrk\ remains one of the most luminous and the most
powerful known at $z<0.1$ \citep[e.g.,][]{SuEtal1998}.

An association of \mrk\ with BAL~QSOs was suggested from 
the start \citep{ArDiEs1971}, but confirmation of this status required ultraviolet 
spectroscopy.  
The first published {\it International Ultraviolet Explorer} (\iue) 
spectra showed no obvious \CIV\ or \MgII\ 
emission or absorption, but the signal-to-noise ratio of the data was low \citep{HuNe1987}.
{\it Hubble Space Telescope} (\hst) 
ultraviolet spectra confirmed the BAL~QSO identification of \mrk\  
with the discovery of broad \MgII\ and \FeII\ absorption 
\citep{SmScAlAn1995}.  
Broad absorption lines are found in $\sim10$\% of
 optically selected QSOs, but \MgII\ BAL QSOs only comprise $\sim10$\% 
of that population.
Other spectral characteristics that \mrk\ shares with low-ionization BAL QSOs include
reddened optical and ultraviolet continua \citep[e.g.,][]{BoEtal1977,HuNe1987}, 
strong optical \FeII\ emission \citep[e.g.,][]{AdWe1972,BoEtal1977}, and weak \OIII\ emission
\citep[e.g.,][]{BoMe1992}.

Though the emission of many ultraluminous infrared galaxies appears to be dominated by
energetic starbursts, \mrk\ has been repeatedly identified as an
exception \citep[e.g.,][]{GoJoDoSa1995,KrCoThKr1997,RiEtal1999}, and
many pieces of evidence point toward an accreting black hole as the major power source behind 
the enormous infrared luminosity (though see Downes \& Solomon 1998\nocite{DoSo1998} 
for an alternate view).  The optical spectrum
exhibits broad \Ha\ \citep[${\rm FWHM}=2870$~\kms; e.g.,][]{BoMe1992} and 
asymmetric \Hb\ emission.  The bright nuclear source is unresolved at radio,
infrared, and optical wavelengths
\citep[e.g., Ulvestad, Wrobel, \& Carilli
1999;][]{LaEtal1998,SuSa1999}\nocite{UlWrCa1999}, and a parsec-scale
radio jet with low apparent speeds has been resolved with
Very Long Baseline Array (VLBA) observations \citep{UlWrCa1999}.  
Variability on time scales of years has been observed in the radio
emission \citep{UlEtal1999} and in the Na~D and \HeI\ absorption lines 
\citep{BoMeMoPe1991,KoDiHa1992,FoRiMc1995}.  
The \iras\ color, $F_{25}/F_{60}$
(the ratio of the flux density at 25~$\mu$m to that at 60~$\mu$m),
is characterized as ``warm'' ($>0.25$), $F_{25}/F_{60}=0.29$ \citep[e.g.,][]{LoHuKlCu1988}.  
This implies a non-thermal continuum or warm dust, more likely to result from the 
hard ionizing continuum of an active nucleus.  In addition, the ratio of infrared luminosity
to mass in H$_2$, $L_{\rm IR}/M_{\rm H_2}=225~L_{\odot}~M_{\odot}^{-1}$,
is difficult to explain without a powerful QSO 
contributing the bulk of the infrared luminosity 
\citep*{SaScSo1991}.{\footnote{For reference, the total
\citet{SaScSo1991} sample has a median value of 
$L_{\rm IR}/M_{\rm H_2}\approx26~L_{\odot}~M_{\odot}^{-1}$, and NGC~4418, the
galaxy with the next highest value after \mrk, has $L_{\rm IR}/M_{\rm
H_2}=89~L_{\odot}~M_{\odot}^{-1}$.}}
In \hst\ images, the point-like nucleus has colors inconsistent with a starburst, 
and $M_{B}\sim-21.6$ with no correction for reddening \citep{SuEtal1998}.  From 
the shape of the optical and ultraviolet continua and infrared line ratios, 
the reddening is estimated to be $A_{V}\sim2$~mag 
\citep{BoEtal1977,CuRiLe1984}.  
Correcting for reddening then places \mrk\ at QSO luminosities ($M_{B}\lesssim-24$) 
that can account for the infrared energy \citep[e.g.,][]{SuEtal1998}.

Though the primary power behind the incredible far-infrared luminosity of \mrk\ is 
almost certainly an active nucleus, the galaxy is also undergoing an energetic starburst.  
The $B-R$ colors of optical knots are indicative of
active, early-type star formation \citep{HuNe1987}, and
\hst\ images of the nucleus clearly resolved blue, star-forming knots
 \citep{SuEtal1998}. 
Most dramatically, in the inner kiloparsec, a nuclear ring of active 
star formation with a rate estimated to be $>100$~\msun~yr$^{-1}$ has
been studied with Very Large Array and VLBA observations \citep{CaWrUl1998,TaSiUlCa1999}.
Furthermore, \citet{SaEtal1987} observed CO emission suggestive of 
many spiral galaxies worth of concentrated
molecular gas that has since been mapped with high resolution by \citet{BrSc1996}.   
The optical morphology of \mrk\ is irregular, and the
asymmetries in the structure of the host galaxy are likely
the result of a merger. Tidal tails extending more than $30\arcsec$ to 
the North and South of the galaxy clearly visible in deep images
further support the merger scenario.
The combination of starburst and luminous QSO make \mrk\ a classic example of the 
transition from starburst to QSO in the paradigm outlined by 
Sanders et al. (1988)\nocite{SaEtal1988}.

X-ray studies of ultraluminous infrared galaxies are essential since they provide a 
direct probe not only of the environment of an active nucleus but also of 
the end-products of stellar evolution.  \mrk\ is of additional interest as a BAL QSO.
BAL QSOs are notoriously weak X-ray sources, and few X-ray spectra of
them exist in the literature.  Since the \heao\ and \einstein\ era,
\mrk\ has been identified as anomalously 
X-ray weak given the implied power of its active nucleus 
\citep{EaAr1988,Ri1988}.
\mrk\ was first detected in X-rays with the \rosat\ Position Sensitive Proportional
Counter \citep[PSPC;][]{RiLaRo1996}, and 
extended emission was observed with the High Resolution Imager \citep[HRI;][]{Tu1999}.  
\asca\ observed this source twice, and spectral analysis indicated both starburst and non-thermal
emission in the X-rays \citep{Iw1999,Tu1999}.  With the most recent 100~ks \asca\
observation, \citet{MaRe2000} argued that the hard X-ray spectrum is dominated
by X-rays ``reflected'' from neutral or nearly neutral material, and
they modeled an \feka\ emission line with EW$\approx290$~eV.  
In this model, the obscuration to the nucleus is so severe that no direct
X-rays below $\sim10$~keV escape along the line of sight.  This implies an absorption
column density
$\gtrsim10^{24}$~\cmsq\ and suggests an intrinsic X-ray luminosity more 
typical of a luminous QSO, although the precise value is highly uncertain.

In this paper, we present a \chandra\ ACIS-S3 observation of \mrk.  
Previous X-ray spectral analysis of this target has been confused by the overlap of the 
starburst and nuclear components due to the large point spread
functions (PSFs) of earlier
X-ray missions. With the 
excellent $1\arcsec$ spatial resolution of \chandra, the nuclear and
disk components can be spatially separated 
for spectral analysis.  At $z=0.042$, $1\arcsec$ is approximately 0.8 kpc
for $H_0=75$~km~s$^{-1}$~Mpc$^{-1}$ and $q_0=\onehalf$. 
The low Galactic column density,
\nh$=(1.03\pm0.10)\times10^{20}$~\cmsq\ \citep*{ElLoWi1989}, 
also makes \mrk\ an excellent target for studying the soft X-rays of
an active starburst galaxy.

%-----------------------------------------------------------------------------------------
\section{Observations and Data Reduction}
\label{sec:obs}

\mrk\ was observed with the  back-illuminated \chandra\ ACIS-S3
detector (G.~P.~Garmire et al., in preparation) 
for 39.8~ks on 2000 October 19. 
The data were processed with the standard \chandra\ X-ray Center (CXC) pipeline 
from which the level~1 events file was used.  The data were filtered on good 
time intervals resulting in a ``live'' exposure of 39.3~ks.  The standard
pipeline processing introduces a $0\farcs5$ software randomization into the
event positions, which degrades the spatial resolution by
$\approx12\%$ \citep{ChaEtal2002}. This randomization is done to
avoid aliasing effects noticeable in observations
with exposure times less than $\sim2$~ks. As spatial resolution was crucial for our
analysis and our observation was significantly longer, we removed this
randomization by recalculating the event positions in sky
coordinates using the {\textsf{\sc CIAO}} 2.0 tool, \textsf{acis\_process\_events}. 
The data were filtered to keep
``good'' \asca\ grades (grades 0, 2, 3, 4, and 6).  Filtering on
``status'' removed $\sim30$ X-ray events from the nucleus because X-rays 
were inappropriately tagged as cosmic ray afterglows.  Therefore, the values
in the status column for cosmic ray afterglows were ignored for the final filter on
``good'' status.

In order to evaluate the reliability of our subsequent analysis, we
checked the data for evidence of pile-up.
Pile-up occurs when two or more photons arrive at a pixel during a single, 
3.2~s frame and alters the spectrum and PSF. 
In the raw and processed data, there was no indication of a read-out
streak (due to X-rays from a bright source which arrived as the CCD was being read 
out), which would indicate a count rate sufficient to result in significantly piled-up data. 
However, given the observed count rate of the nucleus (0.039~ct~\persec), this source is 
within the regime where pile-up at the $\sim5\%$ level is a
possibility. At this low level, our statistical errors are likely to be greater
than any effects of pile-up.
Throughout this paper, wherever the analysis could be sensitive to
the effects of pile-up, we carefully test our results.  The complete analysis
of the possible effects of pile-up, which requires detailed spectral
information, is deferred to $\S${\ref{sec:checks}}.

The light curve of the entire CCD was examined for evidence of
variability.  The background did flare by a factor of 2--3 for $\sim2.2$~ks near the end of the
observation.  For spectral analysis of diffuse emission and image
analysis, this time interval was removed leaving a total of $\sim800$ counts 
in this region.  The count rate of the
nuclear source in \mrk\ was high enough, and the source cell region 
small enough, that the increase in background was not significant, 
and so the higher background times were kept for the nuclear light
curve and spectral analysis.  The nuclear source cell contained a total of
$\sim1450$ counts; over
the course of the flare, less than 2 additional background counts over the entire 
ACIS-S3 band would be expected in a circular region with a radius of $2\arcsec$.

\section{Image Analysis}
\label{sec:imaging}

\subsection{Nucleus and Extended Emission}
\label{sec:imaging1}

Mrk~231 is one of the 212 extragalactic sources that define the 
International Celestial Reference Frame, and thus coordinates accurate
to $\lesssim1$~mas are available from Very Large Baseline
Interferometry \citep{MaEtal1998}. The centroid of the nuclear X-ray emission
from Mrk~231 is within $1\arcsec$ of this position; this is within the typical 
absolute astrometric errors of \chandra.

In order to examine qualitatively the spatial structure of the X-ray
emission, the events were filtered into three broad energy bands: full
(0.35--8.0~keV), soft (0.35--2.0~keV), and hard (2.0--8.0~keV).  The
lower bound of 0.35~keV was chosen to avoid a strong feature in the instrumental
background, and above $\sim8$~keV the effective area of the High-Resolution
Mirror Assembly (HRMA) drops while the particle background rises sharply.
Examining the spatial distribution in these bands gives some
indication of the emission mechanisms in different
physical regions.  Generally, thermal, starburst emission is soft
with $kT\lesssim2$~keV while non-thermal, nuclear emission has a harder 
component with emission up to and beyond the highest energies of the ACIS-S3
bandpass.

To bring out subtle features, the X-ray images were each 
adaptively smoothed at the $2.5\sigma$ level using the algorithm of \citet*{EbWhRa2001}.
The detection of soft, extended emission with the \rosat\ HRI \citep{Tu1999} is confirmed,
but the superior resolution of \chandra\ enables additional structure within the galaxy
to be discerned.  Generally, the extended emission is asymmetric, and the
nuclear source is point-like and off-center with respect to the
extended emission (see Figure~{\ref{fig:xray_images}}).
The galaxy X-ray emission is on the same scale as the optical emission
though the tidal tails are not apparent in X-rays.  From a comparison of the soft and hard images
(Figures~\ref{fig:xray_images}b and \ref{fig:xray_images}c), the galaxy emission can be seen to be primarily soft ($E<2.0$~keV), consistent with thermal, starburst
emission.  In the hard band where non-thermal emission would be
expected to dominate, the emission is much more compact, and
the nucleus becomes even more apparent.  At all energies, the nucleus
contributes significantly to the X-ray flux; this is most clearly seen in the surface plots of the
adaptively smoothed soft-band and hard-band images (Figure~{\ref{fig:xray_surface}}).  

Radial surface-brightness profiles were also calculated in order to
examine quantitatively the intensity distribution; the X-ray
centroid of the nuclear emission was used as the origin.  Out to
radii of $\sim30\arcsec$, the galaxy emission in the soft band is above the
background.  However, outside of $\sim18\arcsec$, the hard-band
emission is consistent with the background level (see
Figure~{\ref{fig:rad_profs}}a).  All of the extended hard-band
emission can be attributed to the wings of the PSF of the central source
(see $\S${\ref{sec:checks} for the complete analysis).

The spatial structure of the X-ray emission is not obviously
correlated with optical or radio structures.  In particular, the bright
knots of starburst emission are not apparent in the X-rays
(Figure~{\ref{fig:optical_cont}}).  
However, the X-ray emission enhancement to the Northeast
(see the X-ray contours in Figure~{\ref{fig:optical_cont}}) of the
nucleus is coincident with the
contours of the $G-B$ (green) color from \citet{HuNe1987}.  These authors
speculated that the green emission is likely from nebular \OIII; the structure 
is reminiscent of ionization cones seen in Seyfert~2 galaxies.
To compare different X-ray emission mechanisms with the
multi-wavelength data, we used the adaptive binning code,
{\textsf{AdaptiveBin}} \citep{SaFa2001}, to create a hardness-ratio map.
Such a map indicates the value of the hardness ratio, which is defined as the
ratio of 2.0--8.0~keV to 0.35--2.0~keV counts, at any given
location. The algorithm of \citet{SaFa2001} 
calculates the hardness ratio in each pixel, and then it combines adjacent pixels
until the errors (propagated from Poisson statistics) on the ratio have dropped below a given
fractional tolerance. 
Given our small numbers of counts, we chose a moderate
tolerance of 0.40; i.e., the hardness-ratio error is less than 40$\%$
in each combination of pixels.
The background-subtracted, hardness-ratio map overlaid with full-band
X-ray contours is displayed in Figure~{\ref{fig:hr_map}}. 
The nuclear region spans a noticeably higher range of hardness
ratios, ${\rm HR_{nuc}}\approx0.6$--1.8, than
the extended galaxy (${\rm HR_{gal}}\approx0.1$--0.4) and shows 
obvious small-scale structure.
With the larger number of counts in this region, finer structure can
be distinguished as the pixels are smaller. An inner region directly 
to the West of the nucleus also shows hard X-ray emission (${\rm HR}=1.25\pm0.4$) which
corresponds to the reddest $B-R$ contours from \citet{HuNe1987}.  This
relationship suggests that the hardness is likely the result of
intrinsic absorption that would also redden the optical color.

\subsection{Non-Nuclear Point Sources}
\label{sec:imaging2}

The {\textsf{\sc CIAO}}~2.0 tool, {\textsf{wavdetect}}, which
implements a wavelet source-detection algorithm, was used to search the full,
soft, and hard-band images. Several point sources within $3\arcmin$ of the
X-ray centroid of Mrk~231 were detected, but none (other than the
nucleus) was found within the optical 
extent of the galaxy.  However, visual inspection of the data 
suggested the presence of a point source within $3\arcsec$ of the
nucleus (see Figure~{\ref{fig:zoom}}).  
Based on our experience with the algorithm, {\textsf{wavdetect}} is
not very sensitive to faint point sources on surface-brightness gradients such as
those found near bright point sources.  Therefore, the lack of a
{\textsf{wavdetect}} detection of this putative point source is not unreasonable.

Within a $1\arcsec$-radius source cell of the possible point source, 
23 full-band counts were registered where $\sim10$ are expected from the local
background.  This background was calculated as the mean number of
counts per pixel in an annulus with an inner radius of $1\arcsec$ and
an outer radius of
$2\arcsec$ normalized to the area of the source cell.  There is only a $0.03\%$ chance
of this occurrence due to Poisson fluctuations.
For maximum sensitivity, the entire observation
was used for this measurement; with such a small source cell the
effect of the background flare should be negligible. 
The errors in the source and background measurements were determined using the
Poisson values for $90\%$ confidence tabulated in \cite{Ge1986}; they were then added in
quadrature to calculate the error in the background-subtracted source
counts: $13^{+6.6}_{-4.9}$ counts for a full-band count rate of 
$(3.3^{+1.7}_{-1.3})\times10^{-4}$~ct~\persec. 
Assuming this source has a power-law slope typical of an X-ray
binary, $\Gamma=1.7$, the 0.35--8.0~keV flux, $F_{\rm
X}$, equals $(2.2^{+1.1}_{-0.8})\times10^{-15}$~\flux, though this value is not
sensitive to the assumed model.  At the redshift of \mrk, this
corresponds to a luminosity, $L_{\rm X}=(7.3^{+3.4}_{-2.8})\times10^{39}$~\lumin.

If this is a single point source radiating isotropically, 
then the Eddington limit at this luminosity
suggests a massive compact object, $M=35$--$90~M_{\odot}$.  Such a
source is consistent with observations 
of other starburst galaxies where several point sources
with comparable luminosities and implied masses 
have been detected \citep[e.g.,][]{MakEtal2000}.
Alternatively, an off-nuclear point source with this flux could be a
luminous supernova remnant \citep[e.g.,][]{Sc1995} such as was found in the
Circinus Galaxy \citep{BaEtal2001}.  Given the low numbers of counts,
this could also be a spatially extended area of enhanced X-ray
emission such as a superbubble.

\section{Variability Analysis}
\label{sec:variability}

Nuclear light curves were examined for variability, and the full-band 
count rate did apparently show a subtle decrease over the course of the
observation. To examine this possible variability 
quantitatively, the full-band event arrival times from a
$2\arcsec$-radius source cell were compared with a constant count rate
using the non-parametric Kolmogorov-Smirnov (K-S) test.  
The K-S statistic of 0.036 gave a probability of 0.042 for the data
being consistent with a constant count rate.
In order to investigate possible variability further, the data were
divided into soft and hard bands and tested separately with the K-S
test.  This strategy was motivated by the different emission
mechanisms contributing in different bands; the soft-band emission may 
have a significant starburst component while the hard-band 
emission should be attributed to the active nucleus.
The soft-band light curve was consistent with constant
flux, and a K-S statistic of 0.033 indicated a probability of 0.353
that the data were consistent with the null hypothesis.  In contrast,
the hard-band light curve yielded a K-S statistic of 0.069 giving a
probability of only 0.002 that the data were consistent with a constant
count rate; this result is highly significant.  The binned light curves are
presented in Figure~{\ref{fig:lc}}, and the bin size was chosen to obtain 
at least 50~counts in each bin. The hard-band bin with the lowest flux
contains $55\pm10\%$ of the counts in the bin with the highest flux
corresponding to a flux decrease of $45\pm14\%$ over 20~ks.

To investigate further the energy 
dependence of the variability, we applied a K-S test to the photon
arrival times in different energy bands, starting with 0.35--8.0~keV.
At each step, the lower limit to the energy band was increased, and
the K-S test reapplied.  The results of this analysis are presented in 
Figure~{\ref{fig:ks}}; the energy band showing the most highly
significant variability (with a K-S probability of $4.2\times10^{-4}$) 
was 2.75--8.0~keV.  This energy-dependent
variability can be understood either as energy bands with softer
lower limits suffering from dilution by a soft, constant flux component, or
as a lack of soft flux from the changing component. For example, the
starburst emission, which contributes primarily to the
0.5--2.0~keV emission even within the nuclear source cell
(see \S{\ref{sec:spec}} for further discussion), is generated on
physical scales of parsecs to kiloparsecs.  Because of the larger
scale of this emission, the flux from the starburst would not be expected to vary on
time scales as short as the exposure time of this observation.
In contrast, the small-scale, accretion-generated X-rays 
dominate the power of the nuclear emission 
at energies $\gtrsim2.0$~keV.  Variability on the time scale of
this observation is thus much more likely to arise from this
physically compact region.  In
the hardest bands, the number of counts decreases rapidly as a result of the
smaller effective area of the telescope, thus reducing the sensitivity
of the K-S test to variability.

In order to confirm that low-level pile-up was not causing false variability,
the K-S test was also applied to an annular region with an inner radius
of $0\farcs5$ and outer radius of $3\arcsec$.  Since possible pile-up
at the $\sim5\%$ level will only affect the innermost pixels of a point
source, an annular source region is much less susceptible to pile-up, though it also
contains less than half the counts in each band. As with the circular
source-cell region, the hard band showed significant variability with
a K-S probability of 0.013 of a constant count rate. The soft-band 
light curve was consistent with a constant count rate with a K-S
probability of 0.892.

\section{Spectral Analysis}
\label{sec:spec}

The $1\arcsec$ spatial resolution of \chandra\ allowed us to perform spectroscopy on 
physically distinct regions.  
Throughout the spectral analysis, Galactic absorption was fixed at
$1.03\times10^{20}$~\cmsq\ (see \S{\ref{sec:intro}}).  All errors are for 90\% 
confidence with all fitted parameters taken to be of interest other than
absolute normalization, unless otherwise noted.  
The counts, fluxes, and luminosities of different regions within the
galaxy are listed in
Table~{\ref{tab:flux}, and the spectral fitting results described
within this section are summarized in Table~{\ref{tab:spec}.

\subsection{Nucleus}
\label{sec:spec_nuc}

\subsubsection{Spectrum Extraction}

Analysis of ground-calibration data from the ACIS back-illuminated (BI)
CCDs (which include S3) showed well before launch that these devices exhibit significant
charge transfer inefficiency (CTI).  CTI occurs during both parallel and
serial transfers, but it is mild enough that the
energy resolution is not position-dependent.  However, the gain,
quantum efficiency, and grade distribution change subtly across the device.
In an attempt to mitigate the effects of CTI, \citet{ToBrNoGa2001} have
developed a software tool for correcting CTI through a
forward-modeling algorithm incorporating a detailed physical CCD
model \citep{TowEtal2001}.  This model has been tuned using an extensive library of 
on-orbit calibration data and tested on astrophysical sources.  
CTI-corrected data require redistribution matrix files (rmfs) and
quantum efficiency uniformity (qeu) files
distinct from those distributed by the CXC; these have been made publicly 
available.{\footnote{\textsf{http://www.astro.psu.edu/users/townsley/cti/}}  
CTI correcting ACIS-S3 data allows a single rmf to be used across the whole CCD thus
making analysis of extended sources more tractable.  In addition,
recovering the initial event grades makes grade filtering
uniform across the CCD, and accounting for the charge lost due to
transfers improves the energy resolution of the X-ray spectrum.

After correcting the data for CTI, the spectrum of the nuclear region
was extracted in pulse-invariant (PI) format 
from a $2\arcsec$-radius circular cell.  The
background spectrum was extracted from an annulus centered on the nucleus with an inner
radius of $2\farcs5$ and an outer radius of $4\arcsec$.  The qeu file for
the CTI-corrected data was used to generate an ancillary response file (arf)
using the {\textsf{\sc{CIAO}}}~2.1 tool \textsf{mkarf} which combines 
the effective area of the mirrors with the qeu and quantum efficiency
file.  Using the FTOOL \textsf{grppha}, the data were grouped into bins with at least 15 counts for
$\chi^2$ fitting. The data were fit with the X-ray spectral analysis tool \textsf{XSPEC} 
\citep{Ar1996}. As the quantum efficiency  
calibration below 0.5~keV is uncertain, the counts below this 
value were ignored. The data above 8.0~keV were also ignored for the
reasons described in \S\ref{sec:imaging1}.

\subsubsection{Basic Spectral Modeling}
\label{sec:basic_spec}

The data were first fit with a power-law model.  Though the fit was
statistically acceptable ($\chi^2/\nu=88.2/81$), the residuals showed
systematic effects.  In particular, there were positive residuals
below 1.0~keV and above 5~keV.  Also, the best-fitting photon index, $\Gamma=0.46\pm0.09$,
was unusually hard; the typical photon index for a radio-quiet QSO is
in the range, $\Gamma=1.7$--$2.3$ \citep[e.g.,][]{ReTu2000,GeoEtal2000}.  
As \mrk\ hosts a starburst ring within 
$1\arcsec$ of its nucleus
\citep{BrSc1996,CaWrUl1998,TaSiUlCa1999}, a plausible additional
component to add is a Raymond-Smith thermal plasma \citep{RaSm1977}.  
The plasma abundance was initially
fixed at solar, a reasonable value for an energetic starburst hosting a
luminous QSO.  The temperature was constrained to lie within the bounds 
found for local starbursts by \citet{PtSeYaMu1999}, $kT=0.2$--3.0~keV.
Compared to the simple power law, this model was 
preferred at the $>95\%$ confidence level
according to the $F$-test ($\Delta\chi^2=-7.8$ for two additional fit
parameters). 
For the thermal plasma, the temperature,
$kT=1.1^{+>1.9}_{-0.2}$~keV, although poorly constrained, 
is consistent with that previously found by
\citet{Iw1999}.  However, the photon index remained atypically flat, 
$\Gamma=0.31^{+0.11}_{-0.20}$. Allowing the abundances to vary did
not substantially improve the fit ($\Delta\chi^2=1.1$), and the
best-fitting abundances were poorly constrained and low, 
$Z=0.32^{+0.63}_{-0.32}~Z_{\odot}$.  As the model fitting was not
very sensitive to this parameter and the best-fitting values appear to be
unreliable, the abundances were fixed to the solar values for the remainder of the analysis.

Intrinsic absorption can often masquerade as a flattening of the
power-law photon index, and so intrinsic absorption was added to the
power-law component.  This did not improve the fit ($\chi^2/\nu=80.5/78$),
and the best-fitting intrinsic column density was negligible.  The
photon index remained at an unusually flat value, $\Gamma\approx0.3$.
To push this physical picture further, the photon index was constrained to lie in 
a reasonable range for radio-quiet QSOs, $\Gamma=1.5$--2.5.  The model
fit for $\Gamma$ ``pegged'' at the lower bound of 1.5 with
$\chi^2/\nu=121.2/78$. Two additional absorption models
were also tested, partial-covering and ionized absorption \citep{DoMuMuAr1992,MaZd1995}, with the
photon index constrained as for the previous example.  In both cases,
the fits were statistically worse than the simple power law plus thermal
plasma, $\chi^2/\nu=92.5/77$ and $\chi^2/\nu=103.5/77$, respectively, 
and the photon index value pegged at $\Gamma=1.5$.  In addition, the
models underfit the flux at energies $\gtrsim5$~keV.

\subsubsection{Reflection-Dominated Spectrum and Iron Line}
\label{sec:ref_spec}

Following the results of \citet{MaRe2000}, we next tried fitting a thermal plasma plus
a power law reflected from neutral material.  Reflected spectra are
commonly seen in Seyfert~1 galaxies in addition to the primary
power-law component; the latter contributes the majority of the flux in the \chandra\
band.  The reflected component is physically understood
to arise from the direct X-ray continuum illuminating a Compton-thick
accretion disk or molecular torus. In Compton-thick Seyfert~2
galaxies, which are viewed along a line of sight through the putative torus,
the primary continuum is completely
absorbed.  Therefore, reflection dominates the emission.

The \textsf{XSPEC} reflection model, \textsf{pexrav} \citep{MaZd1995}, was used to
fit the observed data.
Input parameters to this model include the inclination ($i$) and
abundances of the reflecting medium, as well as the photon index and
cutoff energy ($E_{\rm c}$) of the illuminating power-law spectrum.  The photon index was allowed to
vary, but the other parameters were set to the default values in order to reduce 
the number of free parameters: $\cos(i)=0.45$, $Z=Z_{\sun}$, and
$E_{\rm c}=100$~keV.  
The model was statistically acceptable, $\chi^2/\nu=83.7/78$.
The best-fitting value and errors for the reflection scaling factor,
$R=5737^{+>10~000}_{-2637}$, suggest that reflection is dominating the
spectrum with essentially no direct continuum reaching the observer.
Setting $R$ to produce a pure reflection spectrum gives consistent
values for the photon index and the Raymond-Smith temperature.

The best-fitting photon index, $\Gamma=2.52^{+0.28}_{-0.30}$, though
closer to the typical values for radio-quiet QSOs than those given by
the fits to a simple power-law model,
is unusually high.  Based on the \asca\ correlation for radio-quiet
QSOs between $\Gamma$ and the ${\rm FWHM}_{{\rm H}\beta}$, a photon
index of $\sim2.1$ for \mrk\ would be
more typical \citep{ReTu2000}. In addition to the unusually steep photon
index, the lack of strong \feka\ line emission accompanying a reflected
continuum is also striking. Adding an unresolved Fe~K$\alpha$ line with $E=6.4$~keV and
$\sigma=0.01$~keV to the best-fitting model only decreased $\chi^2$ by 0.2.
\citet{MaRe2000} measured the \feka\ equivalent width to be
${\rm EW}=290^{+190}_{-170}$~eV; however, we do not detect
an \feka\ line at 6.4~keV with high statistical significance. 
Our $90\%$ confidence upper limit on the equivalent width (${\rm EW}<188$~eV), within the
errors of the \citet{MaRe2000} value, is much
lower than would be expected with a reflection-dominated model 
\citep[EW~$\sim1$--2~keV; e.g.,][]{MaBrFa1996}.

Our detection of significant variability above 2~keV is inconsistent
with the proposed model of \citet{MaRe2000} for the nucleus where the reflecting medium 
is the torus, and any scattering medium is on even larger size scales (see their Figure~3).  
Both of these structures, capable of redirecting X-rays
from the nucleus into the line of sight, are on 
physical scales too large for the observed short-term variability.
In any case the reflected component would be 
unlikely to have energy-dependent variability.  
No absorption of the reflected component is required; therefore, the reflected
component would be unlikely to only exhibit variability above 2~keV
as it dominates the flux below 2~keV as well.
Since the reflected component is an unlikely source of the observed
variability, adding a third component, in addition to the Raymond-Smith 
plasma and the reflection-dominated continuum, is thus justified. 
For the spectrum of this third component, we chose an absorbed power
law; adding absorption is motivated by the requirement that this component not
dominate the flux at the lowest energies for consistency with the
observed hard-band variability.

For this fit, the photon indices for both the power-law component and the
incident spectrum on the reflecting medium were fixed at $\Gamma=2.1$
according to the correlation of \citet{ReTu2000}.  Given the quality of
the data, this was necessary in order to obtain meaningful constraints 
on the other model parameters.  The free parameters of interest were
then the Raymond-Smith plasma temperature, the intrinsic absorption column density of the
power-law continuum, and the reflection scaling factor of the
reflection component.  This fit gave a value of
$\chi^2/\nu=69.0/77$, a significant improvement
at the $>95\%$ confidence level over all of the previous models.  The
Raymond-Smith temperature, $kT=1.1^{+1.1}_{-0.1}$~keV, was consistent
throughout all of the model fits; the bump in the spectrum at
$\sim1$~keV corresponds to strong Fe~L shell emission expected from a
plasma of this temperature. 
The reflection scaling factor, $R=697^{+427}_{-223}$, remained large.
This factor is very sensitive to the data below 1~keV; a higher normalization for
the Raymond-Smith plasma component or an additional starburst
component at lower temperature would push the reflection scaling
factor to higher values corresponding to a completely
reflection-dominated spectrum. The value for the best-fitting column
density of the power-law component, $N_{\rm H}=(2.1^{+1.3}_{-0.9})\times10^{22}$~\cmsq,
effectively eliminates flux from the power-law component at energies
$\lesssim2$~keV.

In summary, we find that a three-component model provides a
statistically acceptable fit to the observed nuclear spectrum.  This complex
model is comprised of a Raymond-Smith plasma with
$kT\approx1.1$~keV, an unabsorbed reflection-dominated continuum, plus 
an absorbed power law with $N_{\rm
H}=(2.9^{+2.5}_{-1.0})\times10^{22}$~\cmsq.
Though this composite model provided a good match to the observed spectrum (see
Figure~\ref{fig:nuc_final}a), it has three significant physical
problems.  First, the strength of the absorbed power law in this model is not sufficient
to dilute the EW of the strong \feka\
line expected from the reflected component to the observed upper
limit, ${\rm EW}<188$~eV.  In this model, the absorbed power law
contributes $\sim20\%$ of the observed flux at 6.4~keV.  Manually increasing the normalization of
the absorbed power-law component to provide the 
necessary dilution near 6.4~keV results in a continuum spectrum inconsistent
with these data.  Secondly, the absorbed power law does not
provide enough flux above $2$~keV to explain the observed amplitude of
hard-band variability (see Figure~\ref{fig:nuc_final}b).  Finally, the 
variability time scale implies a physical size of $\sim10^{15}$~cm for
the emitting medium, much smaller than any plausible reflecting medium 
such as the BAL wind or a molecular torus.

\subsubsection{Alternate Model}
\label{sec:alternate}

\mrk\ exhibits complex frequency-dependent
optical and ultraviolet polarization that has been attributed to
scattering by dusty clouds \citep[e.g.,][]{GoMi1994}.
The nuclear continuum polarization rises with decreasing wavelength
and may approach values as high as $20\%$ in the near ultraviolet
\citep{SmScAlAn1995}.  The absorption and emission-line polarization
is highly structured, and \citet{SmScAlAn1995} used three separate
scattering media to explain the behavior of the polarization position angle.
Given this evidence that scattering is an important mechanism
contributing to the optical and ultraviolet emission, we considered
that a similar situation may be
present in the X-rays. Following this reasoning, we examined the possibility that the flat
X-ray spectrum is the sum of multiple
scattered power-law components with different absorption column
densities along each line of sight.  This combination could reproduce
the unusually flat X-ray spectrum without requiring a large
contribution from a reflection-dominated continuum.

Assuming $\Gamma=2.1$ as in \S\ref{sec:ref_spec}, a minimum of three separate power
laws, with column densities of
\nh$=10^{21.1}, 10^{22.5}$, and $10^{23.8}$~\cmsq, are required for an
acceptable fit to the data ($\Delta\chi^2/\nu=66.0/75$).
The best-fitting temperature for the Raymond-Smith plasma component
is consistent to within errors with the previous models. 
In this scenario, the second power law dominates the flux between
\mbox{$\sim1.5$--5~keV} (see Figure~\ref{fig:nuc_final}d), the energy regime
contributing most significantly to the observed variability.
This model has a slightly lower value of reduced $\chi^2$ compared
to the reflection-dominated plus
absorbed power-law model, but the improvement is not statistically conclusive.
Though speculative, this model can provide an explanation 
for both the hard-band variability and the lack of a strong \feka\ emission
line if the scattering medium is highly ionized. This model will be
discussed and developed further in $\S$\ref{sec:discussion_var}. 

\subsubsection{A Jet Contribution to the X-ray Flux?}
\label{sec:jet}

For completeness, we have also considered other possible contributions to
the X-ray flux. As mentioned in $\S$\ref{sec:intro}, 
a parsec-scale radio jet was detected in \mrk\ \citep{UlWrCa1999}
with apparent speed, $\beta_{\rm app} \sim 0.14$. The radio flux of the
mas components was found to be variable with a maximum flux density of
60~mJy \citep{UlEtal1999}. 
We investigated the possibility that this small-scale radio jet could contribute
to the X-ray flux in the \chandra\ spectrum by estimating the 
synchrotron-self Compton (SSC) flux expected from the total radio flux 
density (0.1~Jy) of the nucleus. From the jet-to-counterjet ratio
($>45$), \citet{UlWrCa1999} estimate $\beta\gtrsim0.48$ and $\theta\lesssim10\arcdeg$
if relativistic boosting is responsible for the one-sided radio
structure of the source; though these authors favor free-free absorption
as the cause of the one-sided structure, the relativistic boosting
scenario would generate the largest jet contribution to the X-ray flux.
Assuming $\beta=0.48$ and $\theta=10\arcdeg$, the estimated SSC flux
from 0.5--8.0~keV (using eq. 1 from Ghisellini et al. 1993)\nocite{GhPaCeMa1993}
is $\sim 10^{-22}$~\flux, many orders of magnitude lower than the observed \chandra\
flux (see Table~\ref{tab:flux}). Therefore, SSC emission cannot
explain observed X-ray properties.
 
Another possible source of X-rays from the jet is synchrotron emission.  Applying the 
X-ray-to-radio flux density ratios, $F_{\rm 1~keV}/F_{\rm 15~GHz}$, for
the knots in the jet of M87 \citep{BiStHa1991,MarEtal2001}, we obtained the
predicted range of X-ray fluxes for \mrk, 
$F_{\rm 1~keV}=$(0.3--9.4)$\times10^{-5}$~photon~cm$^{-2}$~s$^{-1}$~keV$^{-1}$.
Though this encompasses the X-ray flux density that we
observe, $F_{\rm 1~keV}\approx2\times10^{-5}$~photon~cm$^{-2}$~s$^{-1}$~keV$^{-1}$, the 
steep slope, $\Gamma\sim2.5$ \citep{MarEtal2001}, observed for the knots 
in M87 is inconsistent with these data ($\Gamma\sim0.5$). Therefore, X-ray emission
from the jet is unlikely to contribute significantly to the observed
flux in X-rays.

\subsubsection{Spectral Checks}
\label{sec:checks}

The HRMA preferentially scatters higher energy photons to larger radii.  
Since the spectrum of the nucleus is so hard, our analysis was also
performed on a spectrum extracted from a $3\arcsec$ radius
aperture with background from an annulus with an inner radius of $3\farcs5$ and an 
outer radius of $5\arcsec$.  This source aperture captures more than $90\%$ of the
encircled energy at 6.4~keV.  The results from fitting the nuclear emission
were consistent with those above, though the soft, thermal emission was not as well constrained.
The larger aperture includes more of the extended starburst emission,
and so this result is not surprising as the thermal starburst emission 
from the larger region is unlikely to have a single temperature.

To investigate further the effects of pile-up on our results, the
nuclear spectrum was also fit with the simulator-based fitting 
tool \textsf{LYNX} \citep{ChEtal2000}. 
\textsf{LYNX} utilizes \textsf{XSPEC} \citep{Ar1996} to generate the
incident spectrum, the PSF simulator
\textsf{MARX} \citep{WiHuDa1997} to simulate the mirror response, and
the Penn State ACIS simulator \citep{TowEtal2001} to propagate X-rays
through the detector.  The $\chi^{2}$ statistic between the
simulated and observed spectra is minimized to obtain the best-fitting
input model parameters.  For this analysis, we used model 4 of 
Table~\ref{tab:spec}, but similar results would have been obtained
other statistically acceptable models.  To simulate a spectrum without pile-up, we
propagated the best-fitting incident model through \textsf{LYNX} allowing only one
incident photon per frame.  From this analysis, the average pile-up
fraction for the observation is estimated to be $7.7\%$. 
For a simple power-law fit to our data,
pile-up has the effect of flattening the slope slightly: an incident
spectrum which would be fit with a photon index of $\Gamma=0.4\pm0.07$
without pile-up would be measured to have $\Gamma=0.3\pm0.08$ with
pile-up for our observed count rate.
This level of pile-up would cause $\sim4\%$
error in the flux measurements.  This systematic error is much less
than the statistical errors, and so pile-up has not significantly affected
our spectral analysis.  

The final \textsf{LYNX} output file can also
be used to estimate the extent to which the wings of the PSF of the nucleus would
contaminate the spectrum of the extended galaxy emission (see $\S$\ref{sec:spec_disk}).  
The simulated radial surface brightness profile in the 2--8~keV band is shown in
Figure~{\ref{fig:rad_profs}b}; the nucleus is expected to contribute
approximately 6 counts in the soft band and 16 counts in the hard band
to the annulus between $3\arcsec$ and $25\arcsec$.  These
contributions are negligible.

\subsection{Extended Galaxy Emission}
\label{sec:spec_disk}

From the images presented in $\S$\ref{sec:imaging}, emission
extending $\sim25\arcsec$ from the nucleus is clearly evident.  For the initial spectral
fitting of this emission, a spectrum for the entire galaxy was extracted using
a circular source cell with a $25\arcsec$ radius.  The 
nucleus was excised with a $3\arcsec$ radius circle, and the 
background spectrum was extracted from a region with an inner radius
of $40\arcsec$ and an outer radius of $65\arcsec$.  
The rmf and arf were generated as described
in $\S$\ref{sec:spec_nuc}.

The extended emission is soft, and the majority of the galaxy is not
detected above $\sim2$~keV. Therefore, the data were filtered 
from 0.5--2.0~keV to maintain an adequate signal-to-noise ratio for the
spectral fitting.  This also eliminates any significant contamination
from the nucleus to the spectrum of the extended galaxy (see $\S${\ref{sec:checks}}).  
As a first model, a Raymond-Smith plasma with
solar abundances was fit.  The resulting temperature,
$kT\approx0.4$~keV, was reasonable, but the value of
$\chi^2/\nu=143.7/31$ was clearly unacceptable.  
Substantial positive residuals 
between 1.0--2.0~keV indicated that an additional component was
required, and so a second Raymond-Smith component was added.  The 
new fit was statistically acceptable ($\chi^2/\nu=37.2/29$).
The two temperatures, $kT=0.30^{+0.07}_{-0.05}$~keV and
$kT=1.07^{+0.22}_{-0.18}$~keV, were also within the range typically
seen in starburst galaxies \citep[0.2--3.0~keV; e.g.,][]{PtSeYaMu1999}.  
The spectrum of the galaxy fit with this model is presented in 
Figure~{\ref{fig:ext_final}}.

Some additional systematic 
positive residuals above 1~keV suggested that a power-law component
could also fit the data, and so a power law was substituted for the
second Raymond-Smith plasma component.  The temperature of the
Raymond-Smith plasma increased relative to the cooler one in previous fits, 
$kT=0.80^{+0.07}_{-0.11}$~keV. The best-fitting photon index for the
power law,
$\Gamma=2.6\pm0.7$, was consistent with emission arising from a
population of X-ray binaries, the most likely source
for such a component.  However,  the value of
$\chi^2/\nu=40.6/29$ was not a significant improvement over the previous
model.  

To investigate whether the spectral 
components were spatially distinct, we divided the extended galaxy
into an inner and outer region with roughly equal counts in each annulus.
The inner region had an outer radius of
$10\arcsec$ with the nuclear region excised, and the outer region
excluded the inner region.  The background regions for both were the same 
as for the complete region.

Based on the counts in the spectrum, the data from 0.5--3.0~keV were
kept for the spectral analysis of the inner region.
Briefly, the inner region was fit equally well with either a
two-temperature Raymond-Smith plasma or a power law plus
Raymond-Smith component.  Both fits produced an acceptable value of
$\chi^2/\nu=14/12$ (see Table~\ref{tab:spec}).  The spectrum of the 
outer region showed less hard X-ray emission, and so the data
were filtered as for the entire extended region, $0.5<E<2.0$~keV.  The
Raymond-Smith plasma plus power law
was slightly preferred over the two-temperature Raymond-Smith plasma
($\Delta\chi^2=1.6$ for the same number of degrees of freedom).

\section{Discussion}
\label{sec:discussion}

\subsection{Mrk~231 as a Broad Absorption Line QSO}

\subsubsection{Spectroscopic Classification}

The definitive classification of Mrk~231 as a BAL~QSO has
been debated in the literature for almost three decades.  The
canonical definition of a BAL~QSO, as published in
\citet{WeMoFoHe1991}, is based on a conservative measurement of the equivalent width of the
\CIV\ resonance absorption line system called the balnicity index
(BI).  All QSOs with BI~$>0$~\kms\ are considered to be BAL~QSOs.
Though Mrk~231 is a bright optical source, the large amount of intrinsic
reddening in its spectrum renders it relatively faint in the
ultraviolet, and a high-quality spectrum of the \CIV\ region is not
available in the literature.

A search of the \hst\ archive revealed a Faint Object Spectrograph (FOS) calibration spectrum
of \mrk\ taken on 1996 Nov 21 covering the crucial \CIV\ region.  This
spectrum, displayed
in Figure~\ref{fig:fos_spec} and unpublished to our knowledge,
clearly shows absorption blue-shifted from
the \CIV\ and \NV\ emission lines, as well as possible \AlIII\ and
\SIV\ broad absorption. A rough measure of the balnicity index, 
BI~$\gtrsim600$~\kms, is a quantitative indicator of the BAL~QSO nature
of this object.  In addition, Mrk~231 fits within the BAL~QSO class
based on its X-ray properties.

The quantity \aox, the spectral index of a power law defined
by the flux densities at rest-frame 3000~\AA\ and 2~keV, is a useful parameter for
measuring the X-ray power of a QSO relative to its ultraviolet continuum
emission.  A large, negative \aox\ indicates relatively weak
soft X-ray emission; the mean value of \aox\ for radio-quiet
QSOs is $\approx-1.48$ \citep[e.g.,][]{LaEtal1997} with a typical range from
$-1.7$ to $-1.3$ for objects that do not suffer from X-ray absorption 
\citep*[e.g.,][]{BrLaWi2000}.  Weakness in soft
X-rays is plausibly explained by intrinsic X-ray absorption, which
strongly depresses the observed flux at low-to-moderate energies.  A strong
correlation of large, negative values of \aox\ with the
absorption-line equivalent width of \CIV\ supports this hypothesis
\citep{BrLaWi2000}. As expected, BAL~QSOs populate the extreme
end of this correlation: they are the weakest soft X-ray sources as well
as the QSOs with the most extreme ultraviolet absorption. Recent
spectroscopic observations of three AGN with large, negative \aox\
values and strong \CIV\ absorption, PG~1411+442,
PG~1535+547, and the BAL~QSO PG~2112+059, found direct evidence of
intrinsic X-ray absorption \citep{BrWaMaYu1999,GaEtal2001a}.  

For comparison with the BAL~QSOs of the \citet{BrLaWi2000} sample, we have 
measured the equivalent width of the \CIV\ absorption blueward of 
the expected location of the emission line, EW~$\approx10$~\AA.
Calculating \aox\ to complete the comparison 
is a difficult matter in this complicated object.
Though the rest-frame 2~keV flux density can be determined from the
best-fitting X-ray model, the ultraviolet spectrum suffers from
severe extinction.  A na{\"{\i}}ve measurement of \aox\ from the
observed 3000~\AA\ flux density yields \aox$=-1.72$; however, this
can only be considered an upper bound on the value.  Correcting for
$A_{V}=2$~mag \citep[e.g.,][]{BoEtal1977} using the extinction curves
of \citet{SpFo1992} results in \aox$=-2.08$; this value includes both 
starburst and scattered flux as well as the reddened nuclear continuum 
before the reddening correction and is therefore a reasonable lower limit to \aox.
Regardless of the exact value of \aox, the strong \CIV\ absorption
coupled with weakness in soft X-rays places Mrk~231 at the heavily absorbed
end of the \citet{BrLaWi2000} correlation, which is populated 
primarily by BAL~QSOs.{\footnote{Though \cite{BrLaWi2000} did not 
correct for intrinsic reddening in their ultraviolet measurements,
several of the BAL~QSOs in their sample, e.g., PG~2112+059, do not
have reddened continua.  Those that do, such as PG~1700+518, would
only have more negative values of \aox.}}

\subsubsection{X-ray Variability and Spectral Results}
\label{sec:discussion_var}

With this \chandra\ observation, we can demonstrate convincingly that
the 2--8~keV X-ray luminosity is dominated by the unresolved active nucleus in Mrk~231.
The hard-band radial profile is consistent with the \chandra\ PSF, and 
the significant, hard-band variability provides compelling evidence 
that \chandra\ is probing within light hours of the central black
hole. At the observed rate the flux would decrease by a factor of 2 in $\approx7$~hr.
The $45\pm14\%$ decrease in count rate in the hard band represents a
luminosity difference, $\Delta L_{\rm X}\approx10^{42}$~\lumin;
variability of this magnitude can only reasonably originate in the
immediate vicinity of the active nucleus.
In addition, the compact spatial extent and very hard spectrum of the dominant
energy source preclude a significant contribution by thermal
starburst emission.  As an alternative, a concentrated population of ultraluminous X-ray
binaries cannot reasonably reproduce the variability, the spectrum, or the
luminosity in the 2.0--8.0~keV band, $L_{2-8}=10^{42.2}$~\lumin.

The measured range of \aox, $-2.08<$~\aox~$<-1.72$, suggests a source that is
under-luminous in the X-ray band by a factor of 6--36, and absorption
remains the most likely explanation for this faintness.  
Though the moderate signal-to-noise ratio of these data
is not sufficient to convincingly constrain the multi-component
spectral properties of \mrk, progress has been made. Assuming a
single, direct, highly absorbed continuum as the sole source of the nuclear
X-rays would require a bizarre and contrived absorber in order to force the
model to match the observed flatness of the X-ray spectrum (see $\S$\ref{sec:basic_spec}).
Though a reflection-dominated model such as that examined by
\citet{MaRe2000} remains feasible based on the spectral fitting
alone, our detection of significant variability is inconsistent with
their physical picture.  
The observed X-ray variability during the
\chandra\ observation strongly suggests that continuum emitted on
small scales is contributing significantly to the hard-band flux.
This continuum cannot reasonably be considered to be direct as the
X-ray flux is not sufficient for this luminous QSO, and so we prefer an indirect (i.e.,
reflected or scattered) origin for this variable component. 

For a physically reasonable reflecting medium, the size scale is
too large to account for the short observed time scale of the variability.
The smallest possible source of a reflecting medium 
would be the BAL wind \citep[e.g.,][]{El2000} at distances
typical of the Broad Line Region \citep[e.g.,][]{OglePhD}.  Even
if the BAL wind resides at the smallest radii 
modeled \citep[$R\sim10^{16}$~cm; e.g.,][]{MuChGrVo1995}, however, the 
shortest expected time scale of variability for the reflected continuum 
would be days to weeks as opposed to the hours observed.  Therefore,
even assuming the smallest reasonable scale for reflecting material, a  
reflection-dominated continuum is not favored by our observation of
hard-band variability.

Our alternate model for the nuclear continuum is comprised of three
scattered power-law spectra absorbed by different column densities of
gas, and this model also
provides a statistically acceptable fit to the data.  We interpret these three
power-law components as scattered rather than direct based
on luminosity arguments; after correcting for absorption, none of 
them could provide the 2--10~keV X-ray luminosity of
$L_{\rm X}\gtrsim10^{44}$~\lumin\ expected from the infrared power
\citep[e.g.,][]{RiGiMaSa2000}. 
The short time scale of the variability provides a spatial scale for
at least one of the scattering regions of
$R\lesssim10^{15}$~cm.  An electron mirror Thomson 
scattering X-rays would require densities of $n\gtrsim10^{9}$~\cc\ on these
small scales (assuming a Thomson optical depth of $0.5$).
Given the implied X-ray luminosity, particle density,
and size, the ionization state of this gas can be
estimated: $\xi=L_{\rm X}/(nR^{2})\sim10^{5}$~erg~cm~s$^{-1}$. 

This scenario supplies a coherent explanation for the hard-band variability as well
as the lack of detected \feka\ emission.  The variability above
$\sim2$~keV can be explained by flux changes of a single
component, a scattered power law with intrinsic absorption of \nh$=10^{22.5}$~\cmsq.
Strong \feka\ line emission would not be expected from
scattering off of a highly ionized plasma
($\xi\sim10^{5}$~erg~cm~s$^{-1}$) where the Fe atoms
are largely stripped of electrons \citep{KaBa2001}.

Our lack of detection of any direct component to the top of the
\chandra\ band-pass supports the conclusion of \citet{MaRe2000} that
an intrinsic absorber with a Compton-thick column density of
$\gtrsim10^{24}$~\cmsq\ blocks these X-rays. Furthermore,
for a reasonable geometry, the blockage of the direct emission
combined with the presence of the scattered emission implies that the
Compton-thick absorber is on a roughly similar scale to the scattering 
mirror; absorbers much larger than the mirror would see it as a point
source and would thus block it entirely (see
Figure~\ref{fig:model} for a possible structure).  
Given the small scales implied for this absorber as well
as the requisite column density, this absorber could be identified
with the ``hitchhiking gas'' proposed by \citet{MuChGrVo1995}, which
shields the BAL wind from becoming completely ionized by extreme ultraviolet and soft
X-ray photons.  The hitchhiking gas would not contribute a
reflected continuum because it is so highly ionized, though it could
scatter X-rays into the line of sight.  Thus the hitchhiking gas could 
provide both the Compton-thick absorber along the direct line of sight 
and the small-scale scattering medium that preserves short timescale variability.

For the absorbers of the multiple scattered power laws, the BAL wind
is a reasonable candidate. As mentioned above, radiative acceleration models put the BAL wind on
much larger scales ($\gtrsim10^{16}$~\cmsq) than our proposed scatterer, 
and the column densities for this
wind are expected to be $\sim10^{23}$~\cmsq\ \citep{PrStKa2000}.  This 
column density is similar to the X-ray column densities measured for
BAL~QSOs from X-ray spectroscopy \citep{MaMaGrEl2001,GrEtal2001,GaBrChGa2001b}, as
well as the column densities of two of the three scattered power-law
components in our modeling.  Multi-epoch observations of \mrk\ are
certainly required to investigate the
possibility of multiple scattered components further. These individual absorbed
power-law components might be
expected to vary with time delays dependent on the physical locations of
their scattering media; in this case, flux variations in the different 
energy bands $<1.5$~keV, 1.5--5~keV, and $>5$~keV would provide
additional support for this proposed model.

Based on \rosat\ data, \citet*{BoBrFi1996} and \citet{LaEtal1997} found a significant
correlation between the X-ray photon indices of radio-quiet AGN and the value
for the ${\rm FWHM}_{{\rm H}\beta}$, where those AGN with the
narrowest broad Balmer lines, Narrow-Line Seyfert~1 galaxies (NLS1s),
tended toward the steep end of the
distribution of $\Gamma$.  Subsequent work with \asca\ in the
2--10~keV band has extended this finding to higher energy, although
the spread of $\Gamma$ in the 2--10~keV band is much smaller 
 \citep[e.g., Brandt, Mathur, \& Elvis
1997;][]{ReTu2000}.\nocite{BrMaEl1997} Mrk~231 has relatively narrow
Balmer emission lines, ${\rm FWHM}_{{\rm H}\alpha}=2870$~\kms\
\citep{BoMe1992} and ${\rm FWHM}_{{\rm H}\beta}=3000$~\kms\ \citep{LiTeMa1993},
particularly for a QSO of its luminosity.  
As a low-ionization BAL~QSO, Mrk~231 shares several spectral
characteristics with NLS1s such as weak \OIII\ lines and strong optical
\FeII\ emission.  This similarity suggests a physical link between
the low-ionization BAL~QSOs and the NLS1s; perhaps both populations are
radiating at a higher fraction of the Eddington luminosity \citep[e.g.,][]{BrGa2000}. 
In order to pursue this potential link, a direct measurement of the
underlying photon index 
of a low-ionization BAL~QSO would offer powerful evidence for a connection. 
Unfortunately, with these data and the evident complexity of the X-ray 
spectrum, we are not able to constrain the underlying photon index of the X-ray spectrum.  
The extreme X-ray faintness of other low-ionization BAL~QSOs
\citep[e.g.,][]{GaEtal1999,GrEtal2001} may 
prevent a direct measurement of $\Gamma$ in these objects for many
years.  The complexity we have found in the spectrum of \mrk\ should
be remembered when interpreting X-ray data on more distant (and
therefore fainter and unresolved) low-ionization BAL~QSOs.

\subsection{Mrk~231 as a Starburst Galaxy}

In this observation of Mrk~231, \chandra\ has resolved the extent and
structure of an X-ray luminous starburst.  Over the entire galaxy, the 0.5--2.0~keV luminosity,
$L_{0.5-2.0}$, equals $10^{41.2}$~\lumin, with $\sim30\%$ from the
inner $2\arcsec$ (1.6~kpc). The strong Fe~L shell emission of the
nuclear spectrum provides  evidence for a $\sim1$~keV thermal component 
in addition to the non-thermal spectrum.
This thermal X-ray component supports the radio evidence for star formation in the
subkiloparsec gas disk \citep{CaWrUl1998}. In this disk, the
star-formation rate claimed by \citet{TaSiUlCa1999} based on their
radio observations is 220~\msun~yr$^{-1}$.  Such an active starburst region might
be expected to be more X-ray luminous; M82 has an 0.5--2.0~keV luminosity of
$\sim10^{40.7}$~\lumin\ with a star-formation rate an order of magnitude
lower \citep[e.g.,][]{PtaEtal1997}.  Though its total flux from star 
formation places Mrk~231 at the X-ray luminous end of the starburst
population, it is still about an order
of magnitude fainter from 0.5--2.0~keV than the most X-ray luminous
starburst, NGC~3256 \citep{MoLeHe1999}. The starburst emission
certainly has more than one component, though
discriminating between a two-temperature thermal plasma and a thermal
plasma plus power-law model is not possible at present.  The
temperature of the thermal plasma within the nuclear region is
$\sim1$~keV, and the data are consistent with such a component
contributing at all radii in the galaxy.
However, the best-fitting plasma temperature at a given radius should
only be considered an emission-measure-weighted average as the actual X-ray spectrum is
likely to be complex.  Higher counting statistics and better
low-energy calibration of ACIS-S3 below 0.5~keV 
would contribute significantly to understanding the distribution of
hot gas in this starburst in more detail.

The one possible off-nuclear source is consistent with the
luminous point sources seen in other starburst galaxies.
One source hardly characterizes a population, however, and further
observations would allow a more definitive description of the luminous 
X-ray binary population of this galaxy.

\section{Future Work}
\label{sec:future}

Mrk~231 was the first example of a low-ionization
BAL~QSO with X-ray spectroscopy, and it may be one of very few for
many years as this class populates the X-ray faintest end of 
the BAL~QSO distribution \citep{GrEtal2001}.  From X-ray spectroscopy
of the handful of BAL~QSOs bright enough for such analysis, the typical 
BAL~QSO spectrum appears to be a normal QSO continuum absorbed by gas with column 
densities of $N_{\rm H}=10^{22-23.5}$~\cmsq\ 
\citep[e.g.,][]{GaBrChGa2001b,GaEtal2001a,GrEtal2001}. \mrk\ does not
fit within this paradigm as it appears to have Compton-thick obscuration with 
only indirect X-ray reaching the observer. 
If \mrk\ is typical of the class of low-ionization BAL~QSOs, their extreme faintness may
result from similar spectra.  Spectra such as that exhibited by \mrk,
whether reflection-dominated or composed of several, absorbed power
laws, are much fainter than a direct, absorbed continuum and have a
fundamentally different spectral shape. Though spectroscopy of other objects may be
difficult, a statistical comparison of the hardness ratios of
low-ionization BAL~QSOs and normal BAL~QSOs can investigate this hypothesis.

Furthermore, \mrk\  is likely to be the only host galaxy
of a BAL~QSO imaged in X-rays for many years; BAL~QSOs tend to lie at
much greater distances.  For comparison, IRAS~07598+6508, the next nearest known BAL~QSO,
has a redshift of $z=0.149$.  Even with relatively good quality data
and the advantage of excellent spatial resolution, the X-ray data for
\mrk\ are complex; similar spectra from unresolved sources at higher
redshifts may prove intractable and must be treated with great care.
Additional observations of \mrk\
with \chandra\ to increase the signal-to-noise ratio for both the
extended galaxy 
and nuclear spectra would allow for study of variability on longer
time scales where the amplitude of variability may well be larger and
could correlate with observed changes over time in the BALs.

\mrk\ has been put forth as a possible local example of energetic
starburst/QSO phenomena common in the early Universe.  It
would be interesting to examine the hardness ratio and flux evolution
as a function of redshift of such a galaxy for comparison with sources
found in deep X-ray surveys.  At present however, the spectrum
above 8~keV is highly uncertain, thus making such analysis above $z\approx1$
speculative.   The question of the true power of this QSO remains, and the
far-infrared and hard X-rays are the most likely regimes for
addressing this issue.  There was no OSSE detection of Mrk~231
\citep{DeBlChMc1997}; however, this limit is not very sensitive.  An 80~ks
observation with the Phoswich Detection System (15--300~keV) on \sax\
scheduled for Cycle 5 should provide significant insight into 
the true power in the hard X-ray regime, and further information on whether the
direct continuum is beginning to show through at the hard end of the
ACIS bandpass. 

\acknowledgements
We thank Dan Weedman for sharing his enthusiasm for and knowledge
of this extraordinary galaxy.  Leisa Townsley and Pat Broos generously provided
expertise and software pertaining to ACIS-S3 in general and CTI correction in particular.
We thank Bill Keel for graciously providing optical images of Mrk~231. 
Mike Eracleous, Mike Brotherton, Franz Bauer, and Nahum Arav also contributed advice and
information.  
The insightful comments of an anonymous referee improved  this paper.
NASA grant NAS 8-38252, P.I. G.~P. Garmire, supports the ACIS Instrument Team. 
S.~C. Gallagher gratefully acknowledges support from NASA GSRP grant NGT5-50277 and
from the Pennsylvania Space Grant Consortium. W.~N. Brandt thanks NASA 
LTSA grant NAG5-8107.

%%%%%%%%%%%%%%%%%%%%%%%%%%%%%%%%%%%%%%%%%%%%%%%%%%%%%%%%%%%%%%%%%%%%%%%%%%%%%%%%%%

%%%%%%%%%%%%%%%%%%%%%%%%%%%%%%%%%%%%%%%%%%%%%%%%%%%%%%%%%%%%%%%%%%%%%%%%%%%%%%%%%
%--------------------------------------------------------------------------------------
\begin{deluxetable}{lcccc}
\tablewidth{0pt}
\tablecaption{Counts, Fluxes, and Luminosities
\label{tab:flux}}
\tablehead{
\colhead{} &
\colhead{Source Counts} &
\colhead{Background Counts\tablenotemark{b}} &
\colhead{$F_{0.5-2}/F_{2-8}$\tablenotemark{c}} &
\colhead{$L_{0.5-2}/L_{2-8}$\tablenotemark{d}} \\
\colhead{Region\tablenotemark{a}} &
\multicolumn{2}{c}{(0.5--8.0~keV)} &
\colhead{($10^{-14}$~\flux)} &
\colhead{($10^{40}$~\lumin)} \\
}
\startdata
Nucleus	&$1447\pm38$	& $36\pm4
$	& $5.4^{+1.1}_{-1.2}/45^{+13}_{-16}$	& 17.1/153	\\
Diffuse Emission	&$833\pm29$	& $240\pm15$	&$3.5^{+0.3}_{-0.4}/\cdots$		& 
$12/\cdots$	\\
\hspace{1 em}Inner Region	&$323\pm18$& $36\pm6$&$1.8^{+0.3}_{-0.3}/\cdots$		& $6.2/\cdots$	\\
\hspace{1 em}Outer Region &$510\pm23$& $204\pm14$ & $2.1^{+0.6}_{-1.0}/\cdots$ & $7.3/\cdots$	\\
\enddata
\tablenotetext{a}{Regions are defined in
$\S${\ref{sec:spec}}.}
\tablenotetext{b}{Counts in the background regions, defined in
$\S${\ref{sec:spec}}, have been normalized to the areas of the source
regions.}
\tablenotetext{c}{The fluxes are calculated from the best-fitting
models; see Table~{\ref{tab:spec}}.  Errors are determined from the
$90\%$ confidence limits on the absolute normalization of the dominant component.}
\tablenotetext{d}{The luminosities are calculated using $z=0.042$,
$H_{0}=75$~\kms~Mpc$^{-1}$, and $q_0=\onehalf$.}
\end{deluxetable}
\clearpage
%%%%%%%%%%%%%%%%%%%%%%%%%%%%%%%%%%%%%%%%%%%%%%%%%%%%%%%%%%%%%%%%%%%%%%%%%%%%%%%%%
%--------------------------------------------------------------------------------------
\begin{deluxetable}{lccccc}
\rotate
\scriptsize
\tablewidth{0pt}
\tablecaption{Parameters for Model Fits to the \emph{Chandra} Data for Mrk~231\tablenotemark{a}
\label{tab:spec}}
\tablehead{
\colhead{Model\tablenotemark{b}} &
\colhead{$\Gamma$} & 
\colhead{Parameter} &
\colhead{Values} &
\colhead{$\chi^2/\nu$} &
\colhead{$P(\chi^2|\nu)$\tablenotemark{c}} 
}
\startdata
Nucleus\\
1. Power law & $0.46\pm0.09$ & $\cdots$ & $\cdots$ & 88.2/81 & 0.37\\ 
\\
2. Power law and Raymond-Smith plasma\tablenotemark{d} & $0.31^{+0.11}_{-0.20}$ &
$kT$~(keV) & $1.1^{+>1.9}_{-0.2}$ & 80.4/79 & 0.44 \\
\\
3. Reflected power law  & $2.52^{+0.28}_{-0.30}$&
$R$ & $5737^{+>10^4}_{-2637}$  & 83.7/78 & 0.31 \\
~~~~and Raymond-Smith plasma & & $kT$~(keV) & $1.3^{+>1.7}_{-0.6}$ &&\\
\\
4. Reflected power law, & $2.1$\tablenotemark{e} & 
$R$ & $697^{+427}_{-223}$  & 69.0/77 & 0.73 \\
~~~~absorbed power law, and &  $2.1$\tablenotemark{e} & 
$N_{\rm H}$~($10^{22}$~cm$^{-2}$) & $2.1^{+1.3}_{-0.9}$  & \\
~~~~Raymond-Smith plasma & & $kT$~(keV) & $1.1^{+1.1}_{-0.1}$ &&\\
\\
5. Three absorbed power laws and & $2.1$\tablenotemark{e} & 
$N_{\rm H}$~($10^{22}$~cm$^{-2}$) & $0.12^{+0.12}_{-0.09}$  &  66.0/75   & 0.76 \\
 &  $2.1$\tablenotemark{e} & 
$N_{\rm H}$~($10^{22}$~cm$^{-2}$) & $3.21^{+1.43}_{-1.37}$  & \\
 &  $2.1$\tablenotemark{e} & 
$N_{\rm H}$~($10^{22}$~cm$^{-2}$) & $59.6^{+89.0}_{-33.3}$  & \\
~~~~Raymond-Smith plasma & & $kT$~(keV) & $1.34^{+>1.67}_{-0.32}$ &&\\
\\
Extended Emission, $E<2.0$~keV\\
1. Raymond-Smith plasma                &   &$kT$~(keV)& 0.4$^{\rm e}$ & 143.7/31 & $<10^{-6}$ \\
2. Two-temperature Raymond-Smith plasma&   &$kT$~(keV)& $0.30^{+0.07}_{-0.05}$ &37.2/29 & 0.14\\
                                       &   &$kT$~(keV)& $1.07^{+0.22}_{-0.12}$ \\
3. Raymond-Smith plasma and power law  &$2.58^{+0.66}_{-0.55}$ & $kT$~(keV)& $0.80^{+0.07}_{-0.11}$ &40.6/29 & 0.07\\
\\
Inner Region, $E<3.0$~keV\\
1. Two-temperature Raymond-Smith plasma &   &$kT$~(keV)& $0.34^{+0.08}_{-0.10}$& 13.9/12 & $0.31$\\
                                        &   &$kT$~(keV)& $1.21^{+0.41}_{-0.21}$ \\
2. Raymond-Smith plasma and power law   &$2.45^{+0.74}_{-0.59}$ &$kT$~(keV)& $0.84^{+0.20}_{-0.05}$ & 13.6/12 & 0.32\\
\\
Outer Region, $E<2.0$~keV\\
1. Two-temperature Raymond-Smith plasma &   &$kT$~(keV)& $0.26^{+0.12}_{-0.11}$& 11.5/14 & 0.65\\
                                        &   &$kT$~(keV)& $0.98^{+0.25}_{-0.23}$ \\
2. Raymond-Smith plasma and power law   &$3.81^{+2.36}_{-1.12}$ &$kT$~(keV)& $0.81^{+0.14}_{-0.15}$ & 9.9/14 & 0.77\\
\enddata
\tablenotetext{a}{All errors are for $90\%$ confidence taking all parameters to be of interest
other than absolute normalization.}
\tablenotetext{b}{All models have fixed Galactic absorption of 
$1.03\times10^{20}$~\cmsq; see \S{\ref{sec:intro}}}
\tablenotetext{c}{The probability, if the given model were correct, that this value of 
$\chi^2$ or greater would be obtained where $\nu$ is the number of degrees of freedom.}
\tablenotetext{d}{The temperatures for all Raymond-Smith plasmas were
constrained to lie within the range $kT=0.2$--3.0~keV; see \S\ref{sec:spec_nuc}.}
\tablenotetext{e}{The power-law photon index was fixed at $\Gamma=2.1$; see \S\ref{sec:spec_nuc}}
\end{deluxetable}
\clearpage
%%%%%%%%%%%%%%%%%%%%%%%%%%%%%%%%%%%%%%%%%%%%%%%%%%%%%%%%%%%%%%%%%%%%%%%%%%%%%%%%%

\begin{figure*}[t]
\figurenum{1}
%\centerline{\includegraphics[scale=0.9]{f1a.ps}}
\protect\caption{
X-ray images of Mrk~231 adaptively smoothed at the $2.5\sigma$ level displayed on a logarithmic 
scale.  The images have been filtered to eliminate the background
flare leaving an effective exposure of 37.1~ks.  
The white cross marks the precise radio coordinates; the X-ray
centroids in each band are within $1\arcsec$ of this position.
The contours indicate the $2.5\sigma$ significance contours.
{\bf (a)} Full-band (0.35--8.0~keV) image.
{\bf (b)} Soft-band (0.35--2.0~keV) image.
{\bf (c)} Hard-band (2.0--8.0~keV) image.  
\label{fig:xray_images}
}
\end{figure*}
%\clearpage

%\begin{figure*}[t]
%\figurenum{1b}
%\centerline{\includegraphics[scale=0.9]{f1b.ps}}
%\end{figure*}
%\clearpage

%\begin{figure*}[t]
%\figurenum{1c}
%\centerline{\includegraphics[scale=0.9]{f1c.ps}}
%\end{figure*}
%\clearpage

\begin{figure*}[t] 
\figurenum{2}
%\centerline{\includegraphics[scale=1.0]{f2a.ps}}
\protect\caption{
Surface plots of the adaptively smoothed images of Mrk~231 in the 
{\bf(a)} soft-band (0.35--2.0~keV) and {\bf (b)} hard-band (2.0--8.0~keV).  The
nuclear region clearly contributes significantly to the emission at all energies, and the
comparable peak intensity in each band indicates the unusually hard nature
of the point source.  Note the extended emission apparent in the soft
band that is not seen in the hard band.
\label{fig:xray_surface}
}
\end{figure*}
%\clearpage

%\begin{figure*}[t]
%\figurenum{2b}
%\centerline{\includegraphics[scale=1.0]{f2b.ps}}
%\end{figure*}
%\clearpage

\begin{figure*}[t]
\figurenum{3}
\centerline{\includegraphics[scale=0.70]{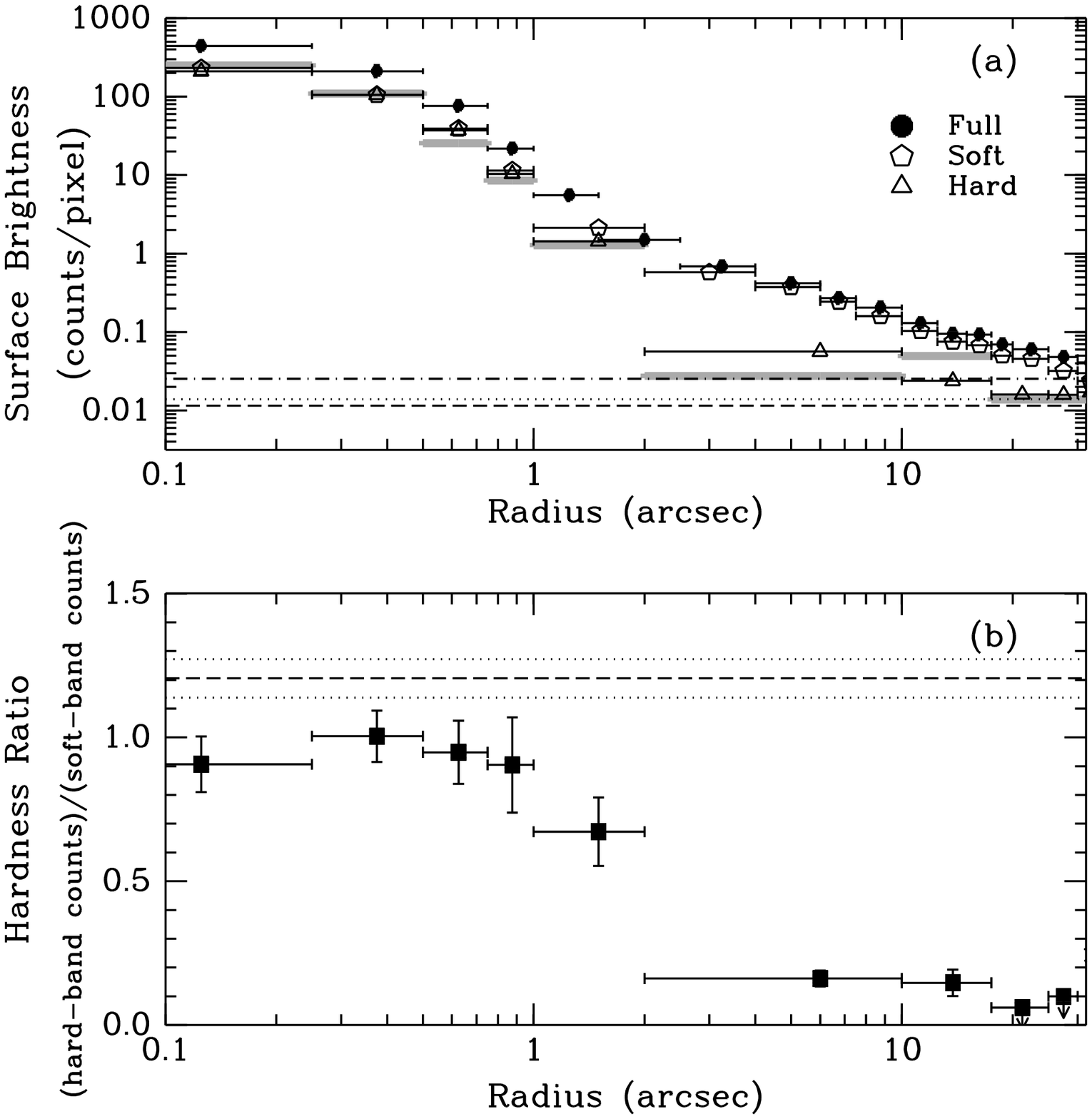}}
\protect\caption{
{\bf (a)} Radial surface-brightness profiles of the emission from
Mrk~231.  Each data point is the counts per pixel in an
annulus with the average radius marked; the annulus width is indicated by
the horizontal error bars. The bin size was chosen so each bin contained 
at least 50~counts, and 
the errors in the vertical direction are comparable to the size of 
the symbols.
The filled circles indicate the surface brightness in the
full-band image while open pentagons and open triangles show the
surface brightness in the soft and hard bands, respectively.  
The thick, gray line is the simulated surface-brightness profile for
the hard band based on
modeling with {\textsf{MARX}} and {\textsf{LYNX}} (see $\S${\ref{sec:checks}}).
The hard-band emission out to $18\arcsec$ is consistent with the PSF.
The dot-dashed, dashed, and dotted lines indicate the surface
brightness of the background for the full, soft, and hard bands,
respectively.
{\bf (b)} Background-subtracted, hardness-ratio profile as a function
of radius.  The nucleus is significantly harder than the soft
extended emission.  Vertical error bars include the errors in the
hardness ratio as well as the background. Where the hard-band emission 
is consistent with the background, the calculated hardness ratio is given as an upper limit.
The dashed line indicates the hardness ratio of the background outside of the galaxy with
the dotted lines showing the errors for that measurement.
\label{fig:rad_profs}
}
\end{figure*}
%\clearpage

\begin{figure*}[t]
\figurenum{4}
%\centerline{\includegraphics[scale=0.50]{f4a.ps}} 
%\centerline{\includegraphics[scale=0.50]{f4b.ps}} 
\protect\caption{
{\bf (a)} Archival \hst\ WFPC2 image taken with the F439W (blue) filter
\citep{SuEtal1998} overlaid with the contours from the adaptively
smoothed, full-band X-ray image.  The
optical image is displayed with a logarithmic scale, and 
X-ray contours represent factors of two in counts. The extended structure in the
X-rays does not appear to correlate with the crescent-shaped star-forming region
to the South of the nucleus evident in the blue image.  
{\bf (b)} Narrow-band H$\alpha$ image \citep{HaKe1987} overlaid with
the contours from the adaptively smoothed, soft-band X-ray image.  The
optical image is displayed with a logarithmic scale, and the contours
represent factors of two in counts.  Note the region of enhanced
emission to the Northeast of the nucleus evident in both the full and
soft-band X-ray contours.
\label{fig:optical_cont}
}
\end{figure*}
%\clearpage

\begin{figure*}[t]
\figurenum{5}
\centerline{\includegraphics[scale=0.80]{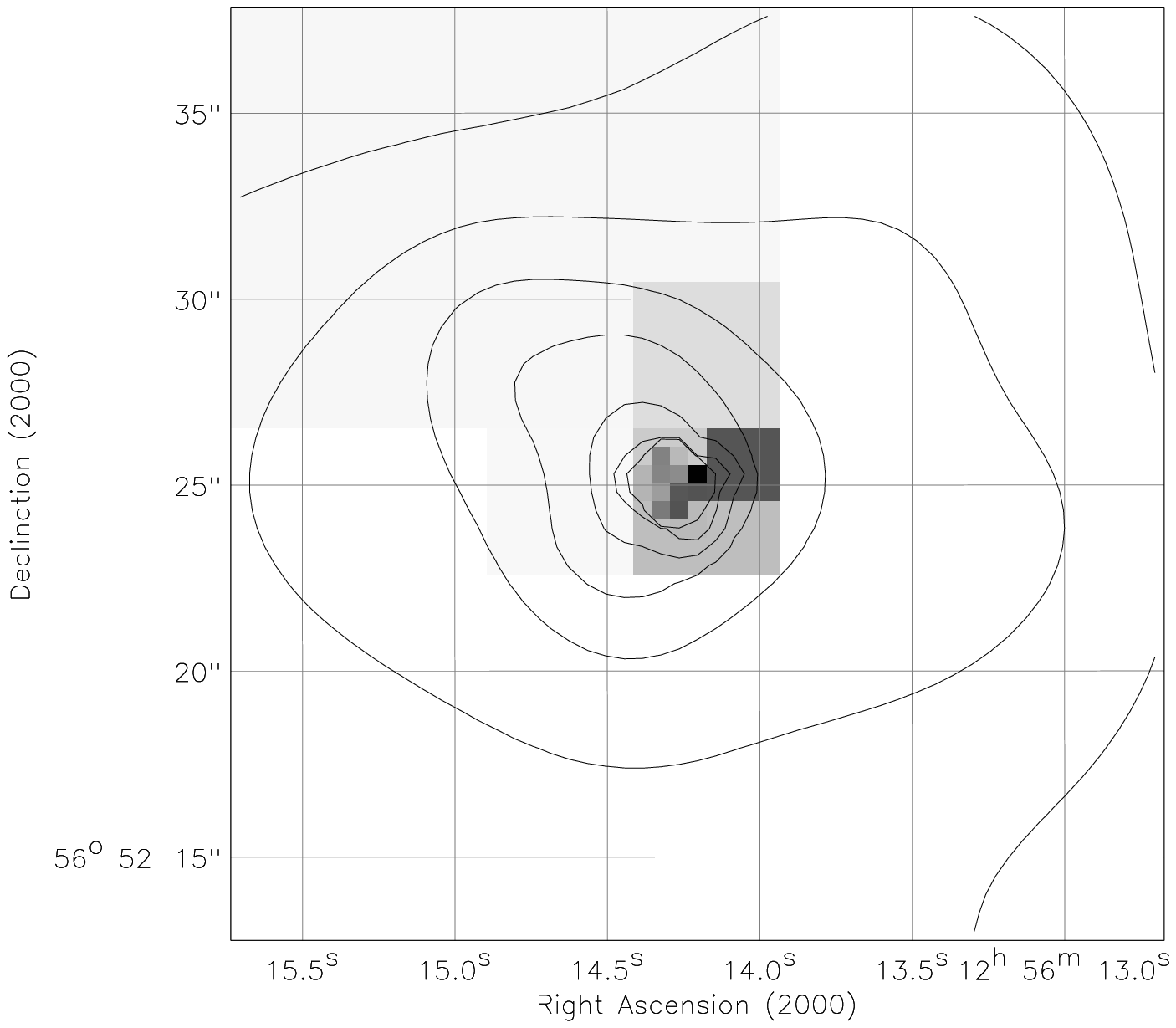}}
\protect\caption{
Adaptively binned X-ray hardness-ratio map showing the ratio of 2.0--8.0~keV to
0.35--2.0~keV counts displayed with a linear scale;
white indicates the softest regions (see $\S$\ref{sec:imaging}).  Only 
the region of the galaxy displayed in this figure has sufficient counts after
background subtraction to measure the hardness ratio.
The nucleus has the highest surface
brightness which results in the finest resolution (or smallest pixels) in 
the map.  The galaxy is significantly softer than
the nucleus; within the errors, the region at radii $\gtrsim2\arcsec$ from the
nucleus has a consistent hardness ratio, HR$_{\rm disk}\approx0.1$--0.2. 
Note the hard region to the West of the
nucleus spatially coincident with the reddest $B-R$ contours from
\citet{HuNe1987}.  The overlaid contours show the intensity from the
adaptively smoothed full-band X-ray image.  Contour levels represent
factors of two in counts.
\label{fig:hr_map}
}
\end{figure*}
%\clearpage

\begin{figure*}[t]
\figurenum{6}
\centerline{\includegraphics[scale=0.80]{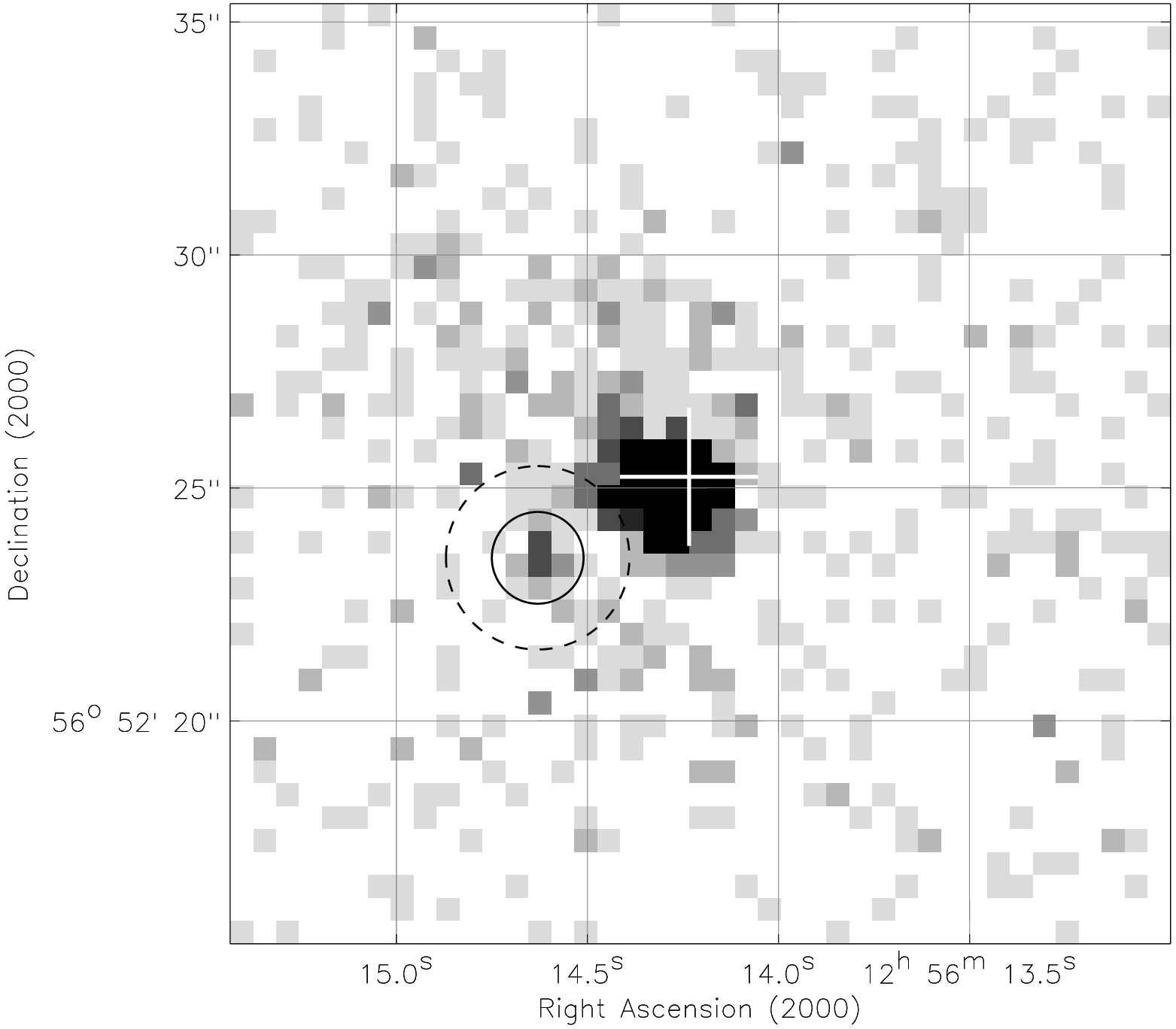}}
\protect\caption{
Full-band image of the $20\arcsec\times20\arcsec$ region immediately
surrounding the nucleus of Mrk~231 displayed with a linear scale.  The
precise radio coordinates of the nucleus are marked with a white cross, and the
putative non-nuclear source, possibly an ultra-luminous X-ray binary or
supernova remnant, is enclosed in the concentric black circles.  
The background region was extracted from the
annulus with an outer radius of $2\arcsec$ (dashed line) and an inner radius 
of $1\arcsec$ (solid line).
\label{fig:zoom}
}
\end{figure*}
%\clearpage

\begin{figure*}[t]
\figurenum{7}
\centerline{\includegraphics[scale=0.80]{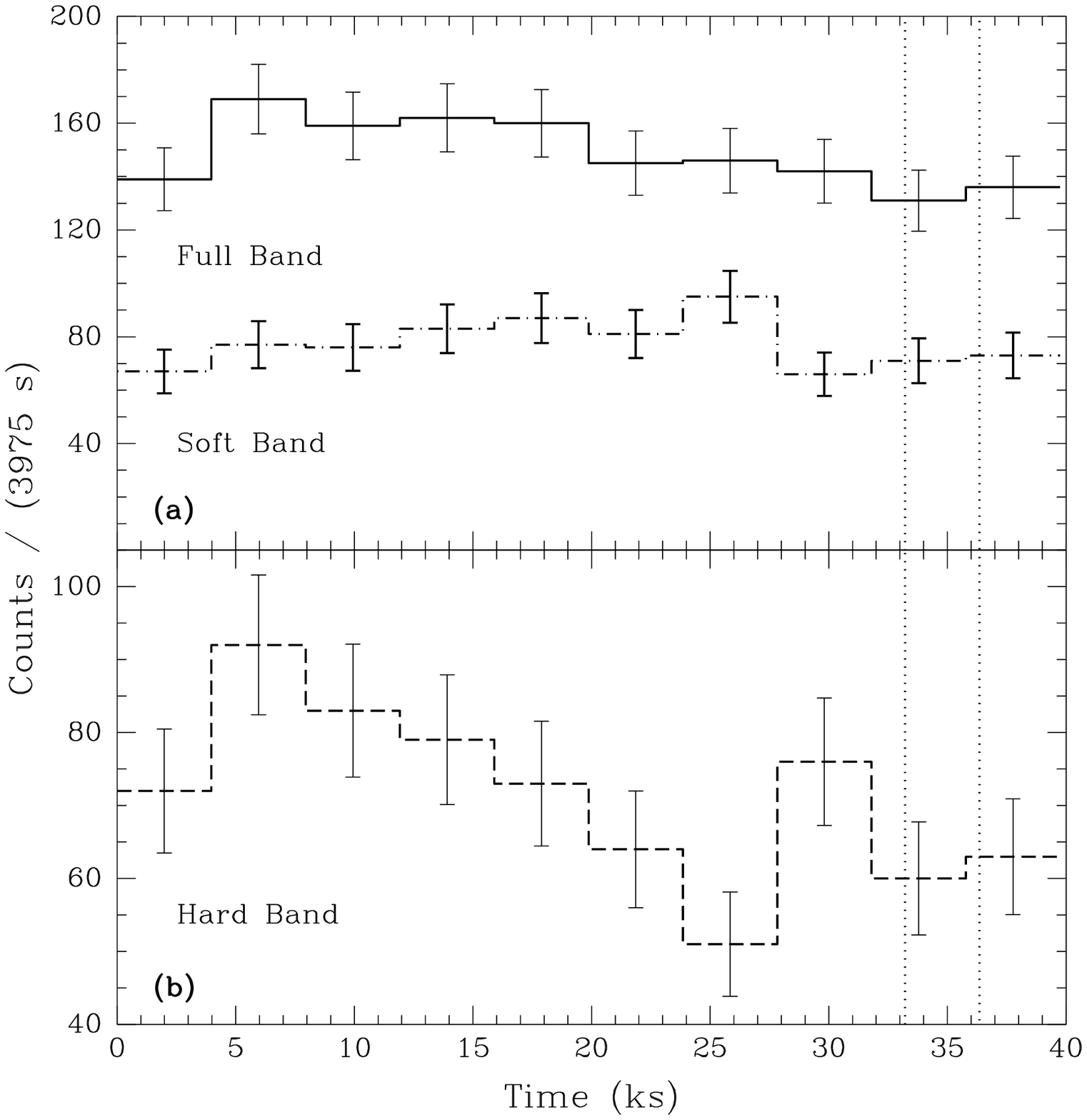}} 
\protect\caption{
Light curves for the nucleus of Mrk~231 extracted from a
circular region with a $2\arcsec$ radius. Each time bin has a width
of 3975~s; the background is expected to contribute $\lesssim1$~count per 
bin.
Error bars are the square root of the number of counts in each bin.
The background flare region (see \S{\ref{sec:obs}}) is indicated with vertical dotted lines.
{\bf{(a)}} Full-band and soft-band light curves shown as solid and dot-dashed 
curves, respectively.  While the full-band light curve shows some
indication of decreasing flux over the course of the observation, 
the soft-band light curve is consistent with a constant count rate.
{\bf{(b)}} Hard-band light curve (dashed line) clearly indicating
a count rate decrease of $\approx45\%$.  Note that the vertical axis does
not extend to zero.
\label{fig:lc}
}
\end{figure*}
%\clearpage

\begin{figure*}[t]
\figurenum{8}
\centerline{\includegraphics[scale=0.80]{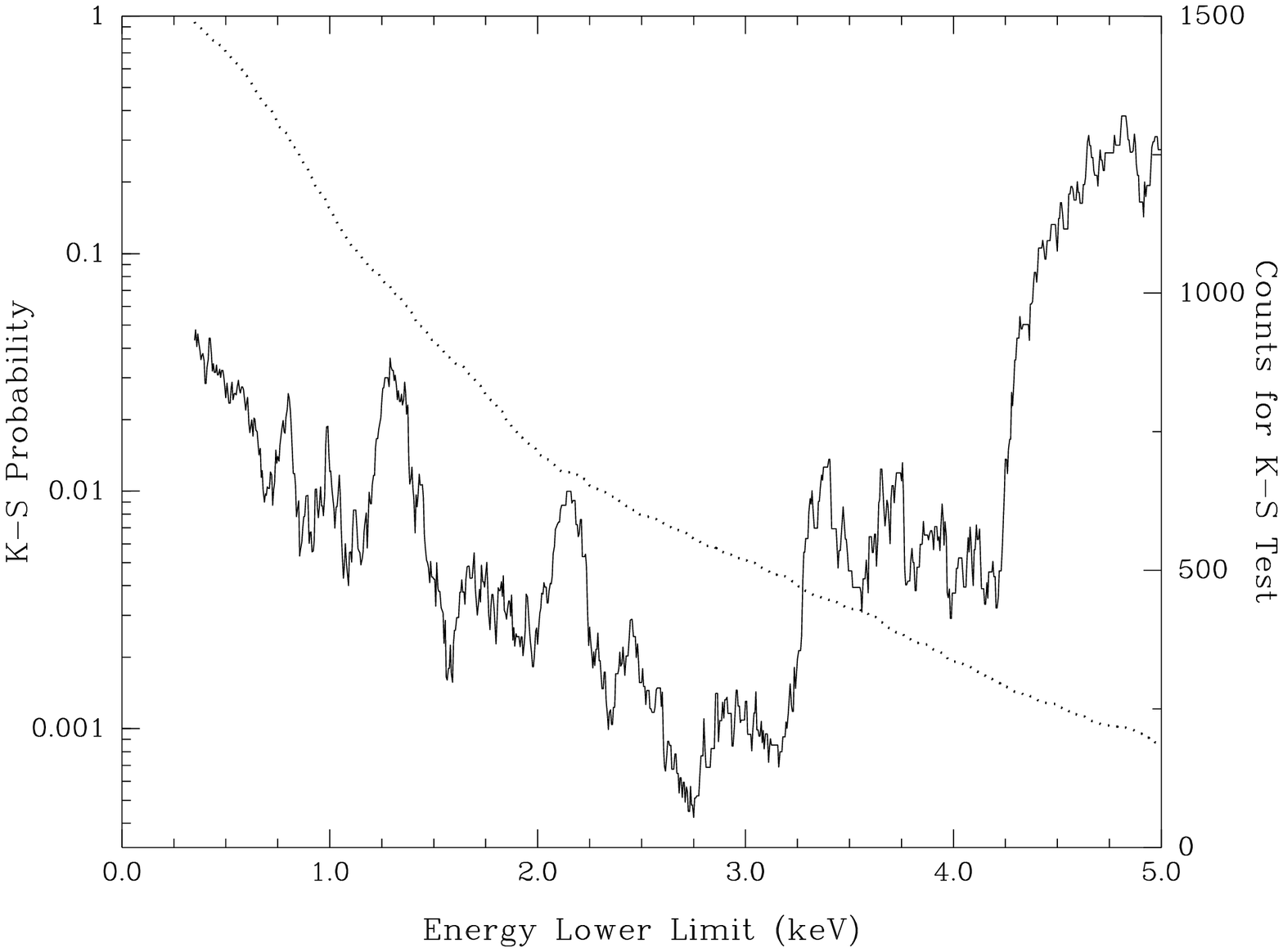}} 
\protect\caption{
The solid curve shows the 
K-S probability as a function of the lower limit on the energy 
band to which the K-S test was applied; all energy bands extend to an upper
bound of 8.0~keV.  
The most significant variability is
indicated by the minimum value of the K-S probability,
$4.2\times10^{-4}$, for an energy band of 2.75--8.0~keV.  
The dotted curve (associated with the right
vertical axis) shows the number of counts used in the K-S test in 
each energy band.  Above  
an energy lower limit of $E\approx 4.25$~keV, 
the calculated significance decreases sharply as the number of 
counts used in the K-S test drops.
\label{fig:ks}
}
\end{figure*}
%\clearpage

\begin{figure*}[t]
\figurenum{9}
\centerline{\includegraphics[scale=0.4,angle=-90]{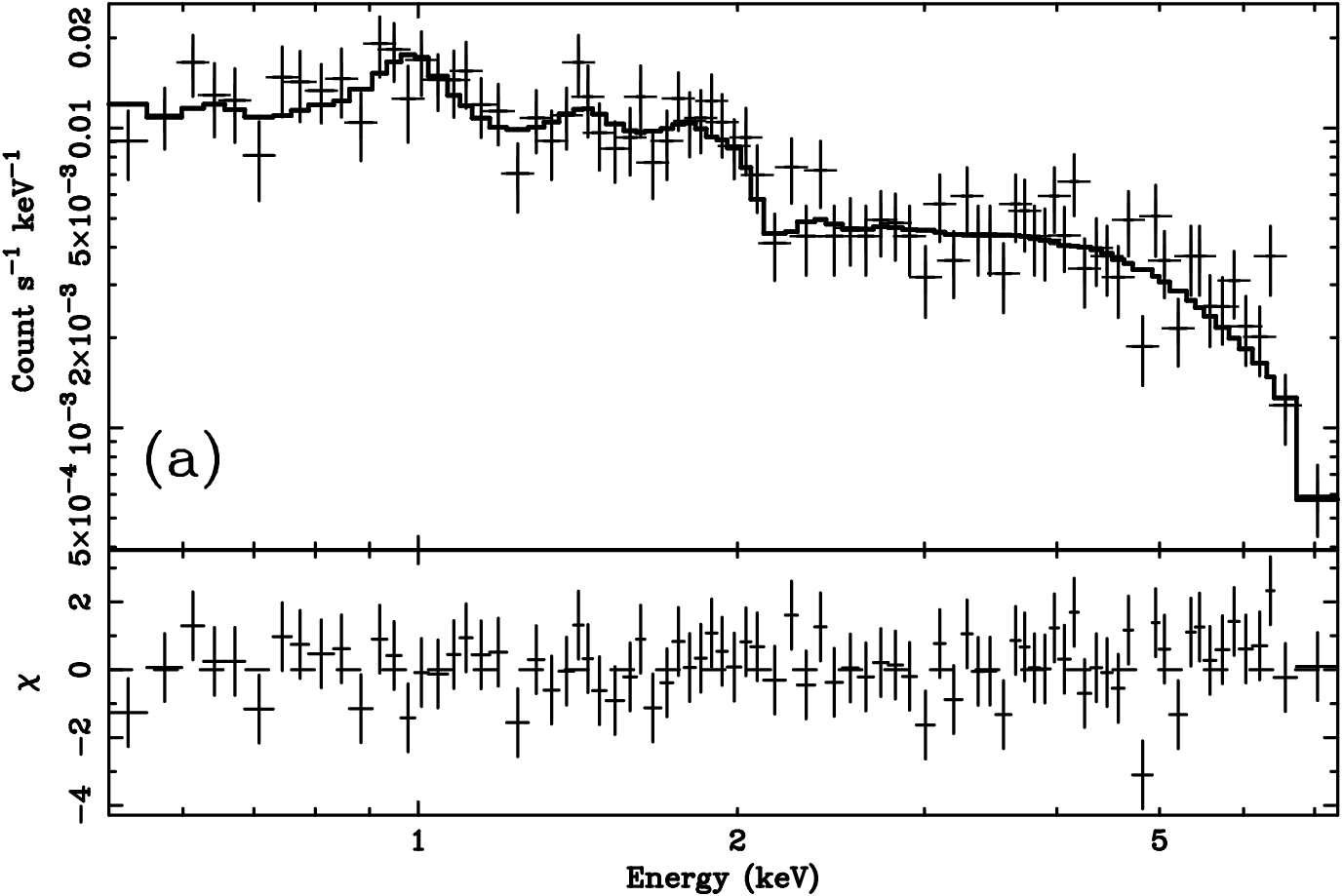}\includegraphics[scale=0.40,angle=-90]{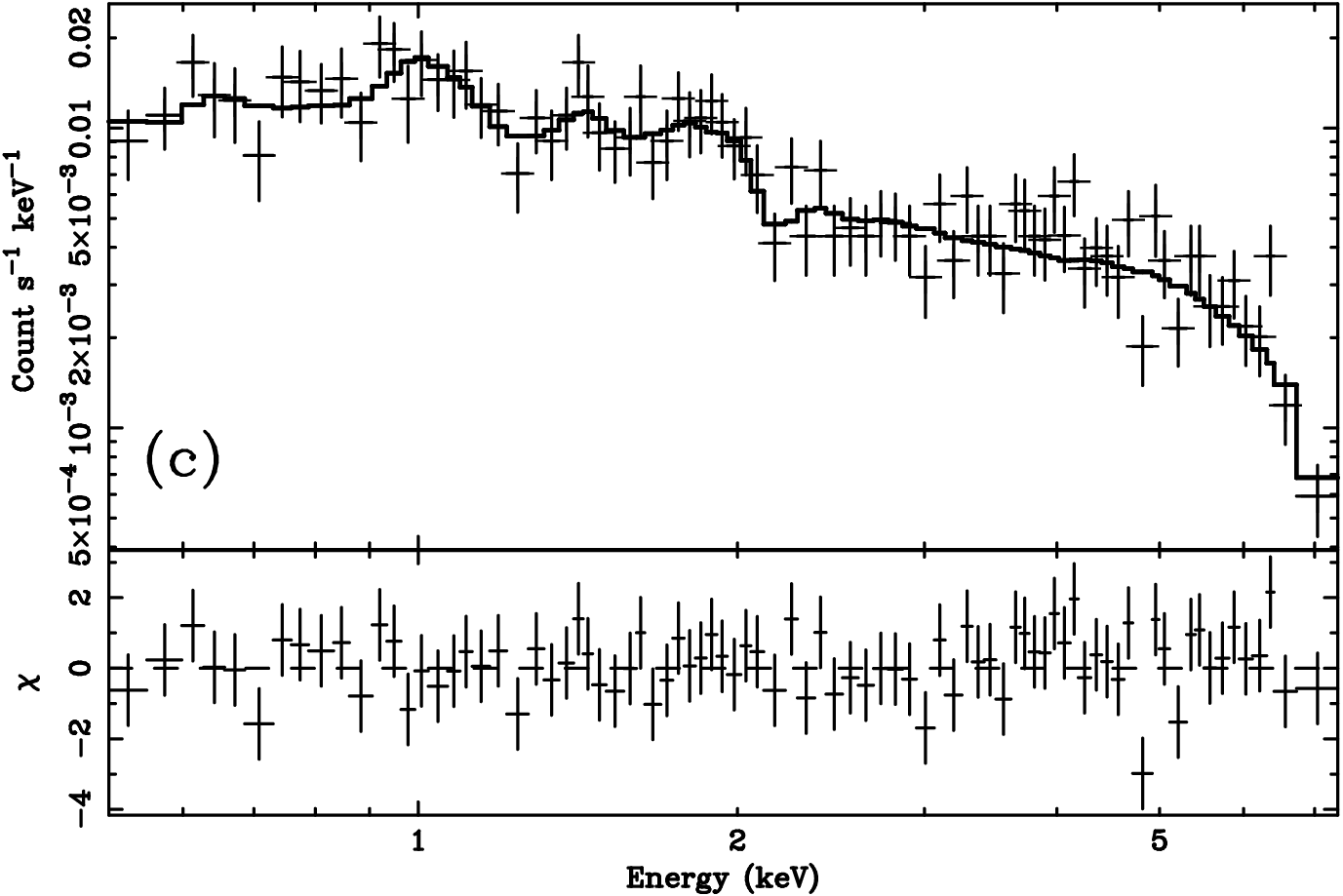}}
\centerline{\includegraphics[scale=0.4,angle=-90]{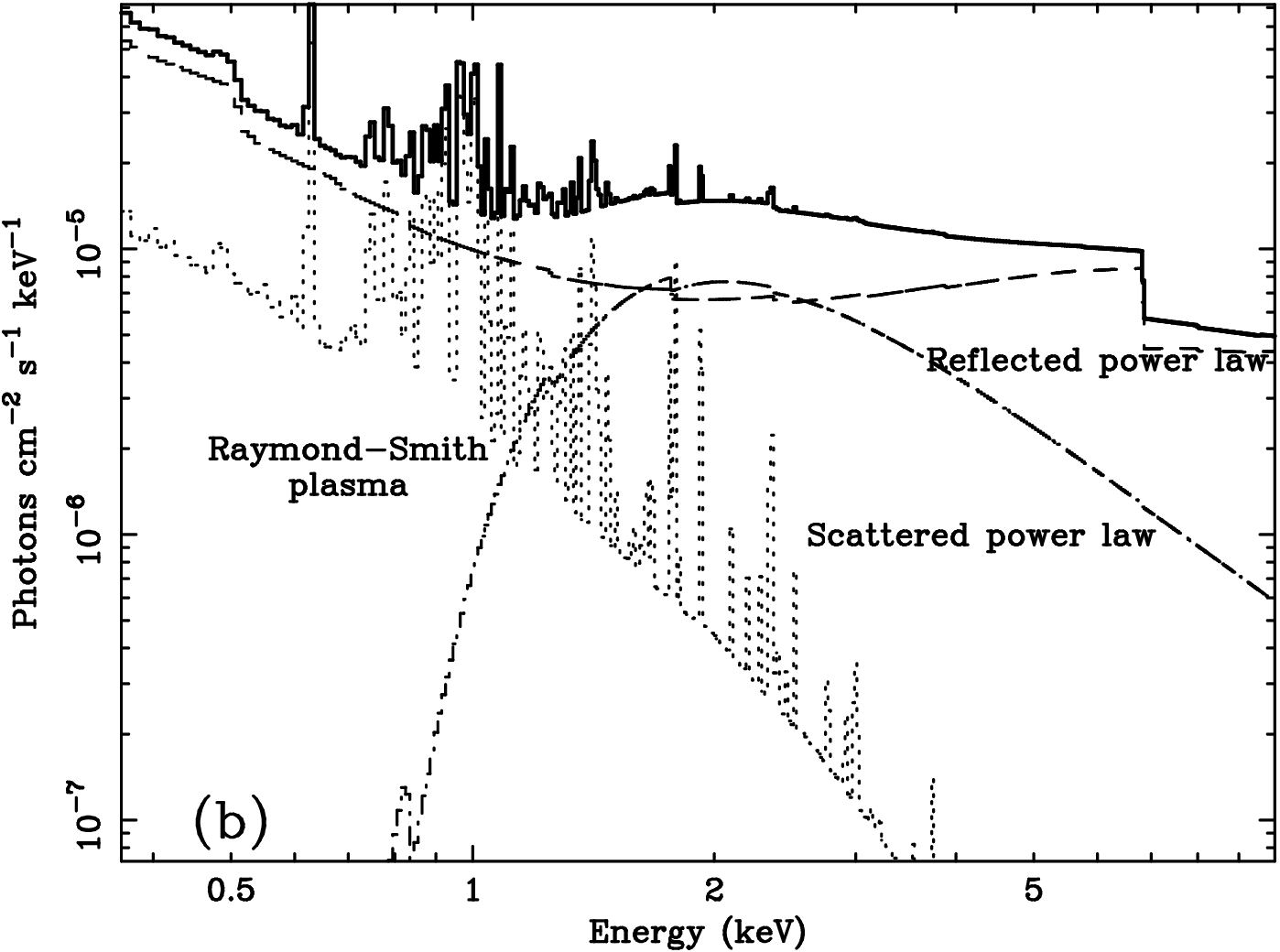}\includegraphics[scale=0.40,angle=-90]{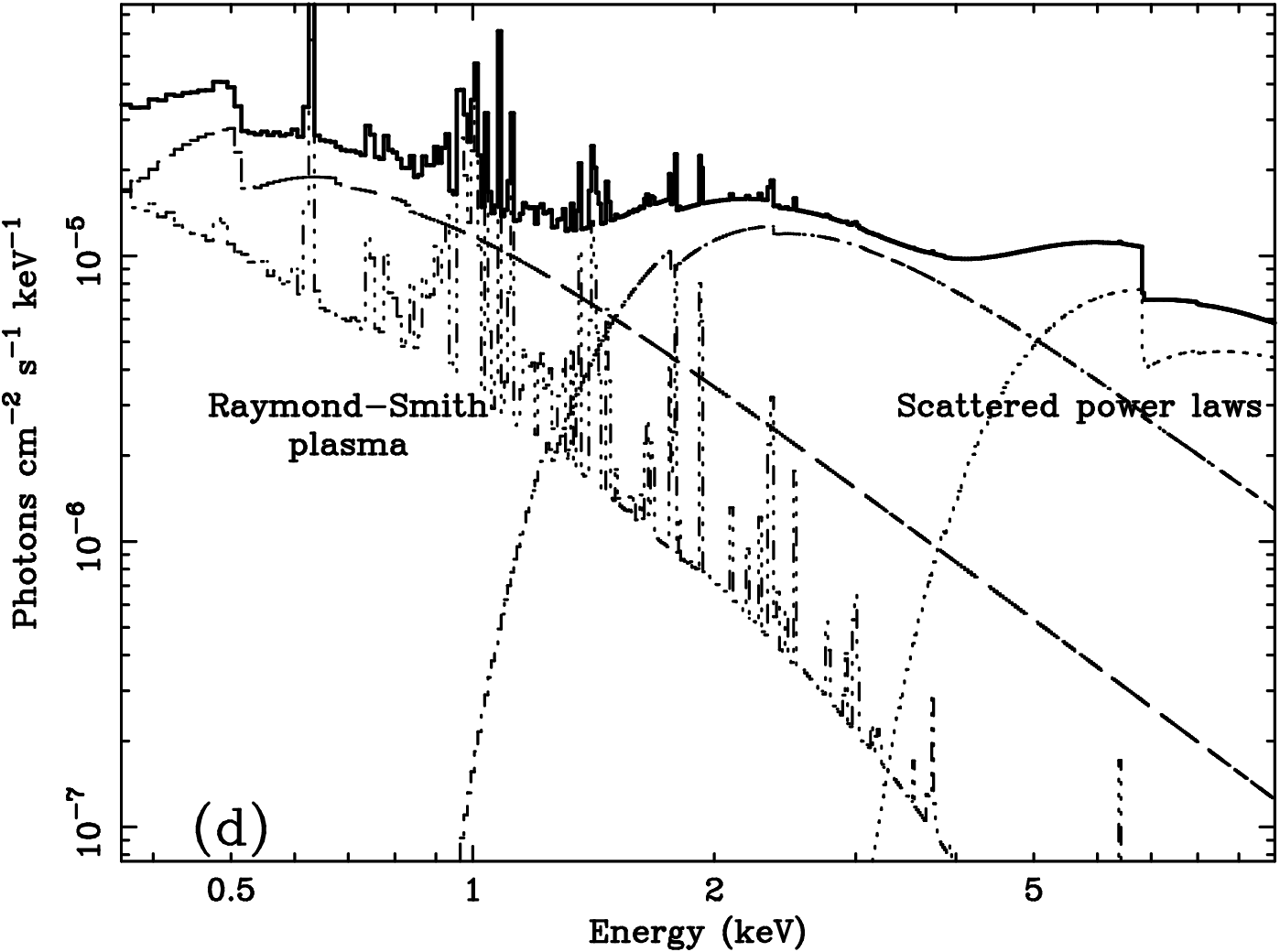}}
\protect\caption{
{\bf (a)} Observed-frame \chandra\ ACIS-S3 spectrum of the nucleus of
\mrk\ fit with a reflected power law, a scattered and absorbed power law, and a
Raymond-Smith plasma (model 4 in Table~{\ref{tab:spec}}).  
In both {\bf (a)} and {\bf (c)}, plain crosses are the data points, and
the solid curve represents the best-fitting model.  The ordinate for
the lower panel, labeled $\chi$, shows the fit residuals in terms of
$\sigma$ with error bars of size one.
{\bf (b)} Best-fitting reflection-dominated model for the nuclear data.
{\bf (c)} Observed-frame \chandra\ ACIS-S3 spectrum of the nucleus of
\mrk\ fit with three absorbed power laws and a Raymond-Smith plasma
(model 5 in Table~{\ref{tab:spec}}).
{\bf (d)} Best-fitting model with multiple absorbed and scattered
power laws for the nuclear data.
\label{fig:nuc_final}
}
\end{figure*}
%\clearpage

\begin{figure*}[t]
\figurenum{10}
\includegraphics[scale=0.70,angle=-90]{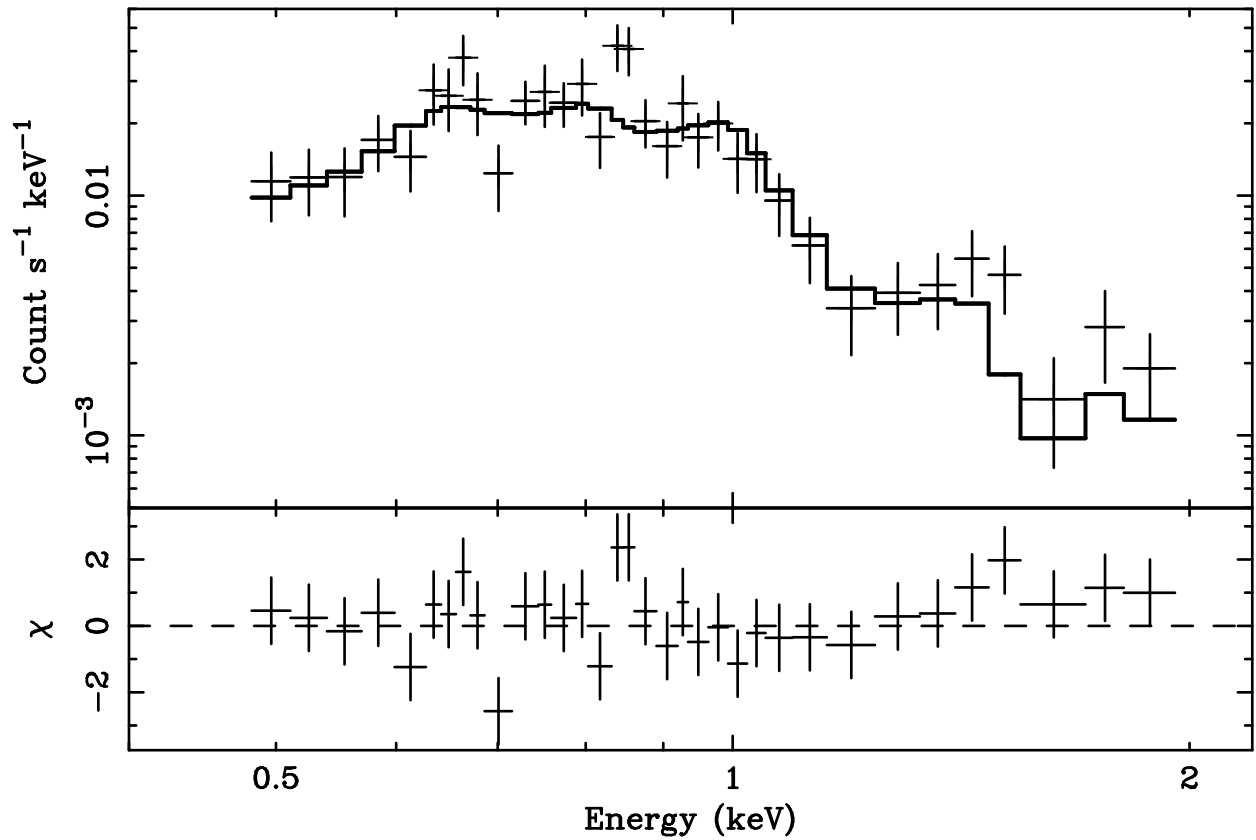}
\protect\caption{
Observed-frame \chandra\ ACIS-S spectrum of the extended galaxy
emission of \mrk\ fit with two Raymond-Smith plasmas (see
Table~{\ref{tab:spec}}).  Plain crosses are the data points, and
the solid curve represents the best-fitting model.  The ordinate for
the lower panel, labeled $\chi$, shows the fit residuals in terms of
$\sigma$ with error bars of size one.
\label{fig:ext_final}
}
\end{figure*}
%\clearpage

\begin{figure*}[t]
\figurenum{11}
\centerline{
\includegraphics[scale=1.00]{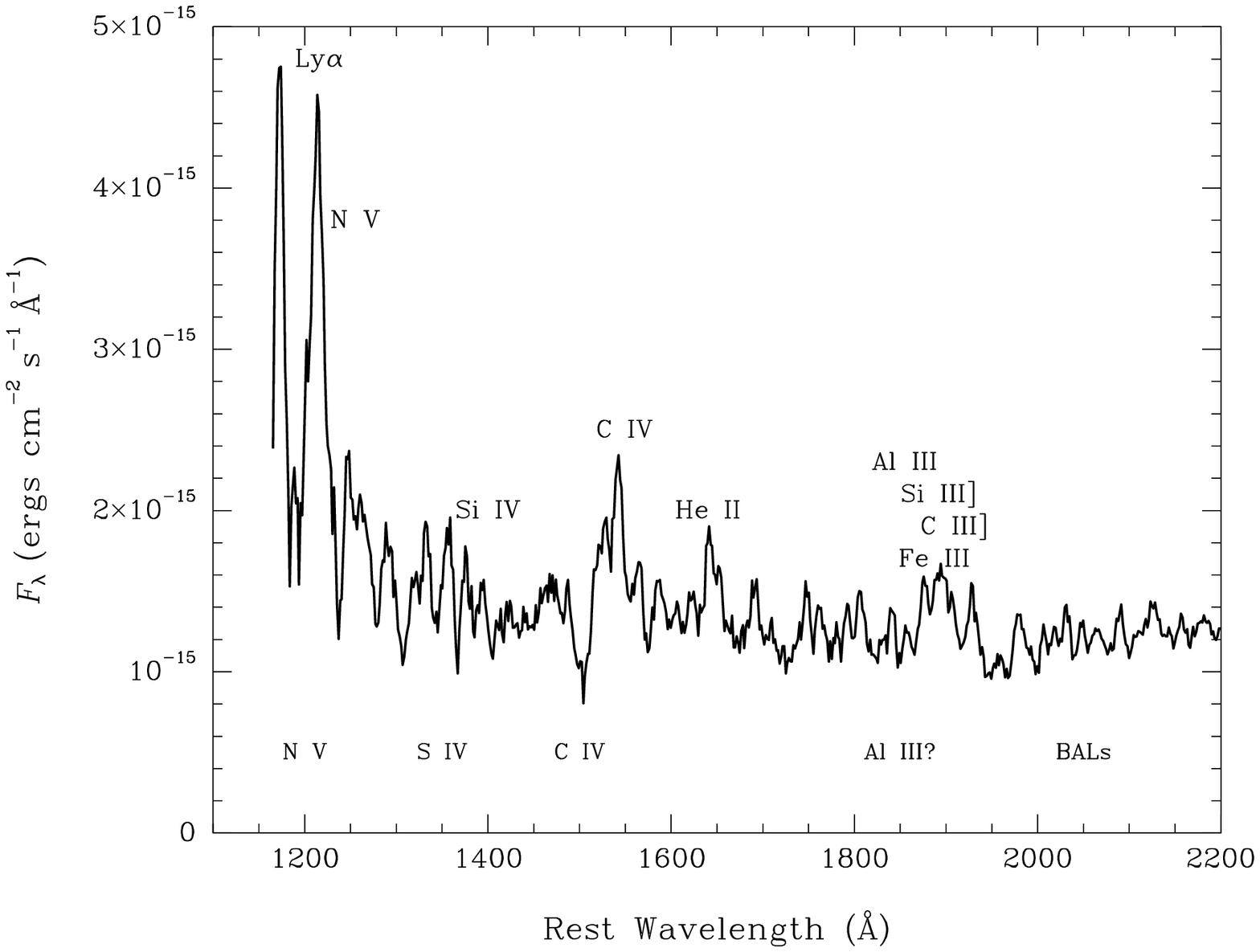}
}
\protect\caption{
Rest-frame \hst\ ultraviolet spectrum of \mrk.  This 770~s FOS calibration
observation was taken with the $1\arcsec$ aperture and the G190L
grating.  The data have been smoothed with a boxcar filter, and they 
have not be corrected for Galactic reddening.
\label{fig:fos_spec}
}
\end{figure*}

%\clearpage
\begin{figure*}[t]
\figurenum{12}
\centerline{
\includegraphics[scale=0.50,angle=0]{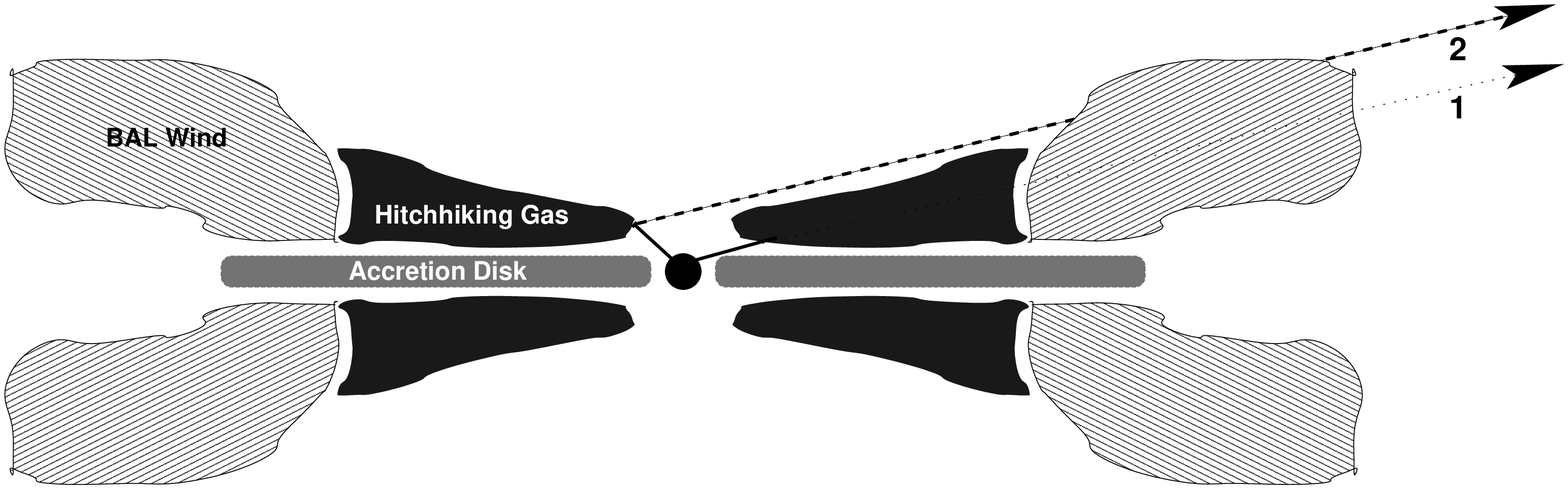}
}
\protect\caption{
Diagram of a possible geometry for the nucleus of \mrk\ (see
$\S$\ref{sec:alternate} and $\S$\ref{sec:discussion_var}).  The filled
black circle represents the black hole and X-ray emitting region.   The
region marked ``Hitchhiking Gas'' has \nh~$\gtrsim10^{24}$~\cmsq, thus
completely blocking direct X-rays at least up to
$\sim8$~keV.  
Two lines of sight are labeled: (1) the direct continuum, which is
completely blocked, 
and (2) the scattered continuum exhibiting the observed X-ray variability.
This second component is observed to be absorbed with
\nh$\sim10^{22}$~\cmsq, perhaps by the BAL wind launched at
distances $\gtrsim10^{16}$~cm.  Note that the
hitchhiking gas would have to be virtually dust free to enable any
optical and ultraviolet continuum to reach the observer.
The other two scattered lines of sight implied by the spectral fitting
($\S$\ref{sec:alternate}) are not shown in this figure, but they would 
have similar paths to (2) (though their scattering regions could be on
larger scales).
\label{fig:model}
}
\end{figure*}
%\clearpage


\begin{thebibliography}{}

\bibitem[\protect\citeauthoryear{{Adams} \& {Weedman}}{{Adams} \&
  {Weedman}}{1972}]{AdWe1972}
{Adams}, T.~F.,  \& {Weedman}, D.~W. 1972, \apjl, 173, L109

\bibitem[\protect\citeauthoryear{{Arakelian}, {Dibay}, \& {Esipov}}{{Arakelian}
  et~al.}{1971}]{ArDiEs1971}
{Arakelian}, M.~A., {Dibay}, E.~A.,  \& {Esipov}, V.~F. 1971, Astrofizika, 7,
  177

\bibitem[\protect\citeauthoryear{Arnaud}{Arnaud}{1996}]{Ar1996}
Arnaud, K.~A. 1996, in {ASP} Conf. Ser. 101, Astronomical Data Analysis
  Software and Systems V, ed. G.~Jacoby \& J.~Barnes (San Francisco: ASP), 17

\bibitem[\protect\citeauthoryear{{Bauer} et~al.}{{Bauer}
  et~al.}{2001}]{BaEtal2001}
{Bauer}, F.~E., {Brandt}, W.~N., {Sambruna}, R.~M., {Chartas}, G., {Garmire},
  G.~P., {Kaspi}, S.,  \& {Netzer}, H. 2001, \aj, 122, 182

\bibitem[\protect\citeauthoryear{{Biretta}, {Stern}, \& {Harris}}{{Biretta}
  et~al.}{1991}]{BiStHa1991}
{Biretta}, J.~A., {Stern}, C.~P.,  \& {Harris}, D.~E. 1991, \aj, 101, 1632

\bibitem[\protect\citeauthoryear{{Boksenberg} et~al.}{{Boksenberg}
  et~al.}{1977}]{BoEtal1977}
{Boksenberg}, A., {Carswell}, R.~F., {Allen}, D.~A., {Fosbury}, R. A.~E.,
  {Penston}, M.~V.,  \& {Sargent}, W. L.~W. 1977, \mnras, 178, 451

\bibitem[\protect\citeauthoryear{{Boller}, {Brandt}, \& {Fink}}{{Boller}
  et~al.}{1996}]{BoBrFi1996}
{Boller}, T., {Brandt}, W.~N.,  \& {Fink}, H. 1996, \aap, 305, 53

\bibitem[\protect\citeauthoryear{{Boroson} \& {Meyers}}{{Boroson} \&
  {Meyers}}{1992}]{BoMe1992}
{Boroson}, T.~A.,  \& {Meyers}, K.~A. 1992, \apj, 397, 442

\bibitem[\protect\citeauthoryear{{Boroson} et~al.}{{Boroson}
  et~al.}{1991}]{BoMeMoPe1991}
{Boroson}, T.~A., {Meyers}, K.~A., {Morris}, S.~L.,  \& {Persson}, S.~E. 1991,
  \apjl, 370, L19

\bibitem[\protect\citeauthoryear{{Brandt} \& {Gallagher}}{{Brandt} \&
  {Gallagher}}{2000}]{BrGa2000}
{Brandt}, W.~N.,  \& {Gallagher}, S.~C. 2000, New Astronomy Review, 44, 461

\bibitem[\protect\citeauthoryear{{Brandt}, {Laor}, \& {Wills}}{{Brandt}
  et~al.}{2000}]{BrLaWi2000}
{Brandt}, W.~N., {Laor}, A.,  \& {Wills}, B.~J. 2000, \apj, 528, 637

\bibitem[\protect\citeauthoryear{{Brandt}, {Mathur}, \& {Elvis}}{{Brandt}
  et~al.}{1997}]{BrMaEl1997}
{Brandt}, W.~N., {Mathur}, S.,  \& {Elvis}, M. 1997, \mnras, 285, L25

\bibitem[\protect\citeauthoryear{{Brinkmann} et~al.}{{Brinkmann}
  et~al.}{1999}]{BrWaMaYu1999}
{Brinkmann}, W., {Wang}, T., {Matsuoka}, M.,  \& {Yuan}, W. 1999, \aap, 345, 43

\bibitem[\protect\citeauthoryear{{Bryant} \& {Scoville}}{{Bryant} \&
  {Scoville}}{1996}]{BrSc1996}
{Bryant}, P.~M.,  \& {Scoville}, N.~Z. 1996, \apj, 457, 678

\bibitem[\protect\citeauthoryear{{Carilli}, {Wrobel}, \& {Ulvestad}}{{Carilli}
  et~al.}{1998}]{CaWrUl1998}
{Carilli}, C.~L., {Wrobel}, J.~M.,  \& {Ulvestad}, J.~S. 1998, \aj, 115, 928

\bibitem[\protect\citeauthoryear{{Chartas} et~al.}{{Chartas}
  et~al.}{2002}]{ChaEtal2002}
{Chartas}, G., {Gupta}, V., {Garmire}, G., {Jones}, C., {Falco}, E.~E.,
  {Shapiro}, I.~I.,  \& {Tavecchio}, F. 2002, \apj, in press (astro-ph/0108277)

\bibitem[\protect\citeauthoryear{{Chartas} et~al.}{{Chartas}
  et~al.}{2000}]{ChEtal2000}
{Chartas}, G., et~al. 2000, \apj, 542, 655

\bibitem[\protect\citeauthoryear{{Cutri}, {Rieke}, \& {Lebofsky}}{{Cutri}
  et~al.}{1984}]{CuRiLe1984}
{Cutri}, R.~M., {Rieke}, G.~H.,  \& {Lebofsky}, M.~J. 1984, \apj, 287, 566

\bibitem[\protect\citeauthoryear{{Dermer} et~al.}{{Dermer}
  et~al.}{1997}]{DeBlChMc1997}
{Dermer}, C.~D., {Bland-Hawthorn}, J., {Chiang}, J.,  \& {McNaron-Brown}, K.
  1997, \apjl, 484, L121

\bibitem[\protect\citeauthoryear{{Done} et~al.}{{Done}
  et~al.}{1992}]{DoMuMuAr1992}
{Done}, C., {Mulchaey}, J.~S., {Mushotzky}, R.~F.,  \& {Arnaud}, K.~A. 1992,
  \apj, 395, 275

\bibitem[\protect\citeauthoryear{{Downes} \& {Solomon}}{{Downes} \&
  {Solomon}}{1998}]{DoSo1998}
{Downes}, D.,  \& {Solomon}, P.~M. 1998, \apj, 507, 615

\bibitem[\protect\citeauthoryear{{Eales} \& {Arnaud}}{{Eales} \&
  {Arnaud}}{1988}]{EaAr1988}
{Eales}, S.~A.,  \& {Arnaud}, K.~A. 1988, \apj, 324, 193

\bibitem[\protect\citeauthoryear{{Ebeling}, {White}, \& {Rangarajan}}{{Ebeling}
  et~al.}{2001}]{EbWhRa2001}
{Ebeling}, H., {White}, D.~A.,  \& {Rangarajan}, F. V.~N. 2001, \mnras,
  submitted

\bibitem[\protect\citeauthoryear{{Elvis}}{{Elvis}}{2000}]{El2000}
{Elvis}, M. 2000, \apj, 545, 63

\bibitem[\protect\citeauthoryear{{Elvis}, {Lockman}, \& {Wilkes}}{{Elvis}
  et~al.}{1989}]{ElLoWi1989}
{Elvis}, M., {Lockman}, F.~J.,  \& {Wilkes}, B.~J. 1989, \aj, 97, 777

\bibitem[\protect\citeauthoryear{{Forster}, {Rich}, \& {McCarthy}}{{Forster}
  et~al.}{1995}]{FoRiMc1995}
{Forster}, K., {Rich}, R.~M.,  \& {McCarthy}, J.~K. 1995, \apj, 450, 74

\bibitem[\protect\citeauthoryear{{Gallagher} et~al.}{{Gallagher}
  et~al.}{2001a}]{GaBrChGa2001b}
{Gallagher}, S.~C., {Brandt}, W.~N., {Chartas}, G.,  \& {Garmire}, G.~P. 2001a,
  \apj, accepted

\bibitem[\protect\citeauthoryear{{Gallagher} et~al.}{{Gallagher}
  et~al.}{2001b}]{GaEtal2001a}
{Gallagher}, S.~C., {Brandt}, W.~N., {Laor}, A., {Elvis}, M., {Mathur}, S.,
  {Wills}, B.~J.,  \& {Iyomoto}, N. 2001b, \apj, 546, 795

\bibitem[\protect\citeauthoryear{{Gallagher} et~al.}{{Gallagher}
  et~al.}{1999}]{GaEtal1999}
{Gallagher}, S.~C., {Brandt}, W.~N., {Sambruna}, R.~M., {Mathur}, S.,  \&
  {Yamasaki}, N. 1999, \apj, 519, 549

\bibitem[\protect\citeauthoryear{{Gehrels}}{{Gehrels}}{1986}]{Ge1986}
{Gehrels}, N. 1986, \apj, 303, 336

\bibitem[\protect\citeauthoryear{{George} et~al.}{{George}
  et~al.}{2000}]{GeoEtal2000}
{George}, I.~M., {Turner}, T.~J., {Yaqoob}, T., {Netzer}, H., {Laor}, A.,
  {Mushotzky}, R.~F., {Nandra}, K.,  \& {Takahashi}, T. 2000, \apj, 531, 52

\bibitem[\protect\citeauthoryear{{Ghisellini} et~al.}{{Ghisellini}
  et~al.}{1993}]{GhPaCeMa1993}
{Ghisellini}, G., {Padovani}, P., {Celotti}, A.,  \& {Maraschi}, L. 1993, \apj,
  407, 65

\bibitem[\protect\citeauthoryear{{Goldader} et~al.}{{Goldader}
  et~al.}{1995}]{GoJoDoSa1995}
{Goldader}, J.~D., {Joseph}, R.~D., {Doyon}, R.,  \& {Sanders}, D.~B. 1995,
  \apj, 444, 97

\bibitem[\protect\citeauthoryear{{Goodrich} \& {Miller}}{{Goodrich} \&
  {Miller}}{1994}]{GoMi1994}
{Goodrich}, R.~W.,  \& {Miller}, J.~S. 1994, \apj, 434, 82

\bibitem[\protect\citeauthoryear{{Green} et~al.}{{Green}
  et~al.}{2001}]{GrEtal2001}
{Green}, P.~J., {Aldcroft}, T.~L., Mathur, S., Wilkes, B.~J.,  \& Elvis, M.
  2001, \apj, in press (astro-ph/0105258)

\bibitem[\protect\citeauthoryear{{Hamilton} \& {Keel}}{{Hamilton} \&
  {Keel}}{1987}]{HaKe1987}
{Hamilton}, D.,  \& {Keel}, W.~C. 1987, \apj, 321, 211

\bibitem[\protect\citeauthoryear{{Hutchings} \& {Neff}}{{Hutchings} \&
  {Neff}}{1987}]{HuNe1987}
{Hutchings}, J.~B.,  \& {Neff}, S.~G. 1987, \aj, 93, 14

\bibitem[\protect\citeauthoryear{{Iwasawa}}{{Iwasawa}}{1999}]{Iw1999}
{Iwasawa}, K. 1999, \mnras, 302, 96

\bibitem[\protect\citeauthoryear{{Kallman} \& {Bautista}}{{Kallman} \&
  {Bautista}}{2001}]{KaBa2001}
{Kallman}, T.,  \& {Bautista}, M. 2001, \apjs, 133, 221

\bibitem[\protect\citeauthoryear{{Kollatschny}, {Dietrich}, \&
  {Hagen}}{{Kollatschny} et~al.}{1992}]{KoDiHa1992}
{Kollatschny}, W., {Dietrich}, M.,  \& {Hagen}, H. 1992, \aap, 264, L5

\bibitem[\protect\citeauthoryear{{Krabbe} et~al.}{{Krabbe}
  et~al.}{1997}]{KrCoThKr1997}
{Krabbe}, A., {Colina}, L., {Thatte}, N.,  \& {Kroker}, H. 1997, \apj, 476, 98

\bibitem[\protect\citeauthoryear{{Lai} et~al.}{{Lai} et~al.}{1998}]{LaEtal1998}
{Lai}, O., {Rouan}, D., {Rigaut}, F., {Arsenault}, R.,  \& {Gendron}, E. 1998,
  \aap, 334, 783

\bibitem[\protect\citeauthoryear{{Laor} et~al.}{{Laor}
  et~al.}{1997}]{LaEtal1997}
{Laor}, A., {Fiore}, F., {Elvis}, M., {Wilkes}, B.~J.,  \& {McDowell}, J.~C.
  1997, \apj, 477, 93

\bibitem[\protect\citeauthoryear{{L{\'{\i}}pari}, {Terlevich}, \&
  {Macchetto}}{{L{\'{\i}}pari} et~al.}{1993}]{LiTeMa1993}
{L{\'{\i}}pari}, S., {Terlevich}, R.,  \& {Macchetto}, F. 1993, \apj, 406, 451

\bibitem[\protect\citeauthoryear{{Low} et~al.}{{Low}
  et~al.}{1988}]{LoHuKlCu1988}
{Low}, F.~J., {Huchra}, J.~P., {Kleinmann}, S.~G.,  \& {Cutri}, R.~M. 1988,
  \apjl, 327, L41

\bibitem[\protect\citeauthoryear{{Ma} et~al.}{{Ma} et~al.}{1998}]{MaEtal1998}
{Ma}, C., et~al. 1998, \aj, 116, 516

\bibitem[\protect\citeauthoryear{{Magdziarz} \& {Zdziarski}}{{Magdziarz} \&
  {Zdziarski}}{1995}]{MaZd1995}
{Magdziarz}, P.,  \& {Zdziarski}, A.~A. 1995, \mnras, 273, 837

\bibitem[\protect\citeauthoryear{{Makishima} et~al.}{{Makishima}
  et~al.}{2000}]{MakEtal2000}
{Makishima}, K., et~al. 2000, \apj, 535, 632

\bibitem[\protect\citeauthoryear{{Maloney} \& {Reynolds}}{{Maloney} \&
  {Reynolds}}{2000}]{MaRe2000}
{Maloney}, P.~R.,  \& {Reynolds}, C.~S. 2000, \apjl, 545, L23

\bibitem[\protect\citeauthoryear{{Markarian}}{{Markarian}}{1969}]{Ma1969}
{Markarian}, B.~E. 1969, Astrofizika, 5, 286

\bibitem[\protect\citeauthoryear{{Marshall} et~al.}{{Marshall}
  et~al.}{2001}]{MarEtal2001}
{Marshall}, H.~L., {Miller}, B.~P., {Davis}, D.~S., {Perlman}, E.~S., {Wise},
  M., {Canizares}, C.~R.,  \& {Harris}, D.~E. 2001, \apj, in press
  (astro-ph/0109160

\bibitem[\protect\citeauthoryear{{Mathur} et~al.}{{Mathur}
  et~al.}{2001}]{MaMaGrEl2001}
{Mathur}, S., {Matt}, G., {Green}, P.~J., {Elvis}, M.,  \& {Singh}, K.~P. 2001,
  \apjl, 551, L13

\bibitem[\protect\citeauthoryear{{Matt}, {Brandt}, \& {Fabian}}{{Matt}
  et~al.}{1996}]{MaBrFa1996}
{Matt}, G., {Brandt}, W.~N.,  \& {Fabian}, A.~C. 1996, \mnras, 280, 823

\bibitem[\protect\citeauthoryear{{Moran}, {Lehnert}, \& {Helfand}}{{Moran}
  et~al.}{1999}]{MoLeHe1999}
{Moran}, E.~C., {Lehnert}, M.~D.,  \& {Helfand}, D.~J. 1999, \apj, 526, 649

\bibitem[\protect\citeauthoryear{{Murray} et~al.}{{Murray}
  et~al.}{1995}]{MuChGrVo1995}
{Murray}, N., {Chiang}, J., {Grossman}, S.~A.,  \& {Voit}, G.~M. 1995, \apj,
  451, 498

\bibitem[\protect\citeauthoryear{{Ogle}}{{Ogle}}{1998}]{OglePhD}
{Ogle}, P.~M. 1998, Ph.D. thesis, California Institute of Technology

\bibitem[\protect\citeauthoryear{{Proga}, {Stone}, \& {Kallman}}{{Proga}
  et~al.}{2000}]{PrStKa2000}
{Proga}, D., {Stone}, J.~M.,  \& {Kallman}, T.~R. 2000, \apj, 543, 686

\bibitem[\protect\citeauthoryear{{Ptak} et~al.}{{Ptak}
  et~al.}{1999}]{PtSeYaMu1999}
{Ptak}, A., {Serlemitsos}, P., {Yaqoob}, T.,  \& {Mushotzky}, R. 1999, \apjs,
  120, 179

\bibitem[\protect\citeauthoryear{{Ptak} et~al.}{{Ptak}
  et~al.}{1997}]{PtaEtal1997}
{Ptak}, A., {Serlemitsos}, P., {Yaqoob}, T., {Mushotzky}, R.,  \& {Tsuru}, T.
  1997, \aj, 113, 1286

\bibitem[\protect\citeauthoryear{{Raymond} \& {Smith}}{{Raymond} \&
  {Smith}}{1977}]{RaSm1977}
{Raymond}, J.~C.,  \& {Smith}, B.~W. 1977, \apjs, 35, 419

\bibitem[\protect\citeauthoryear{{Reeves} \& {Turner}}{{Reeves} \&
  {Turner}}{2000}]{ReTu2000}
{Reeves}, J.~N.,  \& {Turner}, M. J.~L. 2000, \mnras, 316, 234

\bibitem[\protect\citeauthoryear{{Rieke}}{{Rieke}}{1988}]{Ri1988}
{Rieke}, G.~H. 1988, \apjl, 331, L5

\bibitem[\protect\citeauthoryear{{Rieke} \& {Low}}{{Rieke} \&
  {Low}}{1972}]{RiLo1972}
{Rieke}, G.~H.,  \& {Low}, F.~J. 1972, \apjl, 176, L95

\bibitem[\protect\citeauthoryear{{Rieke} \& {Low}}{{Rieke} \&
  {Low}}{1975}]{RiLo1975}
{Rieke}, G.~H.,  \& {Low}, F.~J. 1975, \apjl, 200, L67

\bibitem[\protect\citeauthoryear{{Rigopoulou}, {Lawrence}, \&
  {Rowan-Robinson}}{{Rigopoulou} et~al.}{1996}]{RiLaRo1996}
{Rigopoulou}, D., {Lawrence}, A.,  \& {Rowan-Robinson}, M. 1996, \mnras, 278,
  1049

\bibitem[\protect\citeauthoryear{{Rigopoulou} et~al.}{{Rigopoulou}
  et~al.}{1999}]{RiEtal1999}
{Rigopoulou}, D., {Spoon}, H. W.~W., {Genzel}, R., {Lutz}, D., {Moorwood}, A.
  F.~M.,  \& {Tran}, Q.~D. 1999, \aj, 118, 2625

\bibitem[\protect\citeauthoryear{{Risaliti} et~al.}{{Risaliti}
  et~al.}{2000}]{RiGiMaSa2000}
{Risaliti}, G., {Gilli}, R., {Maiolino}, R.,  \& {Salvati}, M. 2000, \aap, 357,
  13

\bibitem[\protect\citeauthoryear{{Sanders}, {Scoville}, \& {Soifer}}{{Sanders}
  et~al.}{1991}]{SaScSo1991}
{Sanders}, D.~B., {Scoville}, N.~Z.,  \& {Soifer}, B.~T. 1991, \apj, 370, 158

\bibitem[\protect\citeauthoryear{{Sanders} et~al.}{{Sanders}
  et~al.}{1987}]{SaEtal1987}
{Sanders}, D.~B., {Scoville}, N.~Z., {Soifer}, B.~T., {Young}, J.~S.,  \&
  {Danielson}, G.~E. 1987, \apjl, 312, L5

\bibitem[\protect\citeauthoryear{{Sanders} et~al.}{{Sanders}
  et~al.}{1988}]{SaEtal1988}
{Sanders}, D.~B., {Soifer}, B.~T., {Elias}, J.~H., {Neugebauer}, G.,  \&
  {Matthews}, K. 1988, \apjl, 328, L35

\bibitem[\protect\citeauthoryear{{Sanders} \& {Fabian}}{{Sanders} \&
  {Fabian}}{2001}]{SaFa2001}
{Sanders}, J.~S.,  \& {Fabian}, A.~C. 2001, \mnras, 325, 178

\bibitem[\protect\citeauthoryear{{Schlegel}}{{Schlegel}}{1995}]{Sc1995}
{Schlegel}, E.~M. 1995, Reports of Progress in Physics, 58, 1375

\bibitem[\protect\citeauthoryear{{Smith} et~al.}{{Smith}
  et~al.}{1995}]{SmScAlAn1995}
{Smith}, P.~S., {Schmidt}, G.~D., {Allen}, R.~G.,  \& {Angel}, J. R.~P. 1995,
  \apj, 444, 146

\bibitem[\protect\citeauthoryear{{Sprayberry} \& {Foltz}}{{Sprayberry} \&
  {Foltz}}{1992}]{SpFo1992}
{Sprayberry}, D.,  \& {Foltz}, C.~B. 1992, \apj, 390, 39

\bibitem[\protect\citeauthoryear{{Surace} \& {Sanders}}{{Surace} \&
  {Sanders}}{1999}]{SuSa1999}
{Surace}, J.~A.,  \& {Sanders}, D.~B. 1999, \apj, 512, 162

\bibitem[\protect\citeauthoryear{{Surace} et~al.}{{Surace}
  et~al.}{1998}]{SuEtal1998}
{Surace}, J.~A., {Sanders}, D.~B., {Vacca}, W.~D., {Veilleux}, S.,  \&
  {Mazzarella}, J.~M. 1998, \apj, 492, 116

\bibitem[\protect\citeauthoryear{{Taylor} et~al.}{{Taylor}
  et~al.}{1999}]{TaSiUlCa1999}
{Taylor}, G.~B., {Silver}, C.~S., {Ulvestad}, J.~S.,  \& {Carilli}, C.~L. 1999,
  \apj, 519, 185

\bibitem[\protect\citeauthoryear{{Townsley} et~al.}{{Townsley}
  et~al.}{2001a}]{TowEtal2001}
{Townsley}, L.~K., {Broos}, P.~S., {Chartas}, G., {Moskalenko}, E., {Nousek},
  J.~A.,  \& {Pavlov}, G.~G. 2001a, \nim, accepted (astro-ph/0111003)

\bibitem[\protect\citeauthoryear{{Townsley} et~al.}{{Townsley}
  et~al.}{2001b}]{ToBrNoGa2001}
{Townsley}, L.~K., {Broos}, P.~S., {Nousek}, J.~A.,  \& {Garmire}, G.~P. 2001b,
  \nim, accepted (astro-ph/0111031)

\bibitem[\protect\citeauthoryear{{Turner}}{{Turner}}{1999}]{Tu1999}
{Turner}, T.~J. 1999, \apj, 511, 142

\bibitem[\protect\citeauthoryear{{Ulvestad}, {Wrobel}, \& {Carilli}}{{Ulvestad}
  et~al.}{1999}]{UlWrCa1999}
{Ulvestad}, J.~S., {Wrobel}, J.~M.,  \& {Carilli}, C.~L. 1999, \apj, 516, 127

\bibitem[\protect\citeauthoryear{{Ulvestad} et~al.}{{Ulvestad}
  et~al.}{1999}]{UlEtal1999}
{Ulvestad}, J.~S., {Wrobel}, J.~M., {Roy}, A.~L., {Wilson}, A.~S., {Falcke},
  H.,  \& {Krichbaum}, T.~P. 1999, \apjl, 517, L81

\bibitem[\protect\citeauthoryear{{Weedman}}{{Weedman}}{1973}]{We1973}
{Weedman}, D.~W. 1973, \apj, 183, 29

\bibitem[\protect\citeauthoryear{Weymann et~al.}{Weymann
  et~al.}{1991}]{WeMoFoHe1991}
Weymann, R.~J., Morris, S.~L., Foltz, C.~B.,  \& Hewett, P.~C. 1991, \apj, 373,
  23

\bibitem[\protect\citeauthoryear{{Wise}, {Huenemoerder}, \& {Davis}}{{Wise}
  et~al.}{1997}]{WiHuDa1997}
{Wise}, M.~W., {Huenemoerder}, D.~P.,  \& {Davis}, J.~E. 1997, in ASP Conf.
  Ser. 125: Astronomical Data Analysis Software and Systems VI, ed. G.~Hunt \&
  H.~E. Payne (San Francisco: ASP), 477

\end{thebibliography}
\end{document}